\documentclass[fleqn,usenatbib]{mnras}

\usepackage{newtxtext,newtxmath}

\usepackage[T1]{fontenc}
\usepackage[utf8]{inputenc}
\usepackage{ae,aecompl}

\usepackage{graphicx}   
\usepackage{amsmath}    
\usepackage{multirow}
\usepackage[usenames,dvipsnames]{color}

\newcommand{\msun}{M$_{\odot}$}
\newcommand{\Msun}{M$_{\odot}$}
\newcommand{\teff}{T$_{\rm eff}$}

\title[]{On the role of dust and mass loss in the extended main sequence turnoff of star clusters: the case of NGC\,1783}

\author[D'Antona et al.]{ 
F. D'Antona$^{1}$\thanks{E-mail:franca.dantona@gmail.com, francesca.dantona@inaf.it},
F. Dell'Agli$^{1}$,
M. Tailo$^{2}$,
A. P. Milone$^{3,4}$,
P. Ventura$^{1}$,
E. Vesperini$^{5}$, 
G. Cordoni$^{3,4}$,
\newauthor
A. Dotter$^{6}$,
A. F. Marino$^{7,4}$
\\ \\
$^{1}$Istituto Nazionale di Astrofisica - Osservatorio Astronomico di Roma, Via Frascati 33, I-00040 Monteporzio Catone, Roma, Italy\\
$^{2}$Dipartimento di Fisica e Astronomia Augusto Righi, Università degli Studi di Bologna, I-40129 Bologna\\
$^{3}$Dip. Fisica e Astronomia ``Galileo Galilei'', Univ. di Padova, Vicolo dell'Osservatorio 3, I-35122 Padova \\
$^{4}$Istituto Nazionale di Astrofisica - Osservatorio Astronomico di Padova, Vicolo dell'Osservatorio 5, I-35122 Padova, \\
$^{5}$Department of Astronomy, Indiana University, Bloomington, IN 47405, USA\\
$^{6}$Department of Physics and Astronomy, Dartmouth College, Hanover, NH 03755 USA \\
$^{7}$Istituto Nazionale di Astrofisica - Osservatorio Astrofisico di Arcetri, Largo Enrico Fermi, 5, I-50125 Firenze, Italy  \\
}

\date{Accepted 2023 March 16. Received 2023 March 16; in original form 2022 December 27.}

\pubyear{2020-2021}

\begin{document}
\label{firstpage}
\pagerange{\pageref{firstpage}--\pageref{lastpage}}
\maketitle

\begin{abstract}
The Color Magnitude Diagram (CMD) morphology of the `extended' main sequence turnoff (eMSTO) and upper main sequence (MS) of the intermediate age ($\lesssim 2$\,Gyr) Large Magellanic Cloud Cluster NGC\,1783 shows the presence of a small group of UV-dim stars, that, in the ultraviolet Hubble Space Telescope filters, are located at colors on the red side of the typical ``fan'' shape displayed by the eMSTO. We model the UV-dim stars by assuming that some of the stars which would intrinsically be located on the left side of the eMSTO are obscured by a ring of dust due to grain condensation at the periphery of the excretion disc expelled when they spin at the high rotation rates typical of stars in the Be stage. A reasonably low optical depth at 10$\mu$\  is necessary to model the UV-dim group. Introduction of dust in the interpretation of the eMSTO may require a substantial re-evaluation of previous conclusions concerning the role of age and/or rotation spreads in the MC clusters: the entire eMSTO can be populated by dusty stars, and the reddest UV-dim stars simply represents the tail of the distribution with both maximum obscuration and the dust ring seen along the line of sight. The model stars having higher rotational projected velocity ($v \sin$\, i) are predicted to be preferentially redder than the slowly-rotating stars. The mass loss responsible for the  dust may also cause the non-monotonic distribution of stars in the upper main sequence, with two peaks and gaps showing up in the UV CMD.

 \end{abstract}

\section{Introduction}
\label{intro}
The plethora of information obtained through the Hubble Space Telescope (HST) observations of the Magellanic Clouds clusters have expanded our understanding of the formation and evolution of stars in clusters. After finding that  the great majority of the ancient Globular Clusters of the Galaxy host ``multiple populations", it was interesting to discover that Color Magnitude Diagram (CMD) of the much younger (1.5--2\,Gyr) cluster NGC\,1846 in the Large Magellanic Cloud (LMC) displays a double main sequence turnoff \citep{mackey2007}  possibly compatible with the presence of two main populations born $\sim$300\,Myr apart.
An age spread of $\sim$400\,Myr was also necessary to interpret the extended turnoff found in the Small Magellanic Cloud cluster NGC\,419 \citep{glatt2008}. The PhD work by Milone \citep{milone2009} offered a full view of the CMD of 53 LMC clusters, among which about 70\% displayed an intrinsically broadened, or `extended' main sequence turnoff (eMSTO, hereafter sometimes referred to as 'turnoff fan' for its resemblance to a hand fan), suggesting a prolonged star formation epoch.  The distribution of stars in the eMSTO varies from cluster to cluster, and, in a few cases like in NGC\,1846, is characterized by a bimodal distribution suggesting two main star formation events.

The interpretation of the eMSTO in the LMC and SMC clusters followed for many years the idea of multiple star formation epochs (age spread), with the notable exception of  \cite{bastiandemink2009}, who  suggested that stellar rotation may be the cause the eMSTO, as stellar rotation affects the structure of the star and the inclination angle of the star relative to the observer will change the effective temperature, hence the observed colour.
This proposal was debated  \cite[e.g.][]{girardi2011}, with many studies supporting the age spread view \citep{goudfrooij2011,  rubele2013}. 

Some years after, 
\cite{milone2015} identified {\it a split upper main sequence} in the LMC cluster NGC\,1856, in addition to the eMSTO. The complex CMD of  NGC\,1856 was examined by
 \cite{dantona2015}. By using populations synthesis based on the Geneva models, they showed that the cluster data are well interpreted as superposition of two main populations having the same age ($\sim$350\,Myr), composed for 2/3 of very rapidly rotating stars, defining the upper turnoff region and the redder main sequence, and for 1/3 of slowly or non rotating stars. 
Also here, the eMSTO is attributed to the random viewing angles distribution of the very rapidly rotating stars, subject to limb and gravity darkening\footnote{In the Geneva population synthesis, the limb darkening follows \cite{claret2000} and the model for gravity darkening is by \cite{espinosa2011}.}. 

The analysis of another LMC cluster in the same age range, NGC\,1866 ($\sim$200\,Myr) by \cite{milone2017} showed a more complex situation, and that a younger population was also needed to interpret the blue MS, in addition to two main coeval populations with different rotation rates. Although it includes a small percentage of the total cluster population, a  `younger' upper part of the MS is present in several clusters. \cite{dantona2017} showed that it could be attributed to recently `braked' stars, that spend most of their lifetime rapidly rotating and, when braked, appear nuclearly younger than stars born not rotating, because the fast rotating convective core has a larger extent and the MS lifetime is longer. This hypothesis brings back the possibility that the populations in the eMSTO are coeval, and is possibly confirmed by the first spectroscopic determination of rotation velocities in the eMSTO, made by \cite{dupree2017} for NGC\,1866, showing that the blue MS stars are indeed slow rotators as expected if they are braked. A similar scheme ---bluer MSTO stars are mainly slow rotators, red MSTO are mainly rapid rotators--- is found in other spectroscopic data on the eMSTO, see, e.g.,  \cite{marino2018a} for the very young ($\sim$40\,Myr) LMC cluster NGC\,1818 and also for the Galactic cluster M\,11 \citep{marino2018b}. 
%
%
\begin{figure}
\centering
\vskip -40pt
\includegraphics[width=8.5cm]{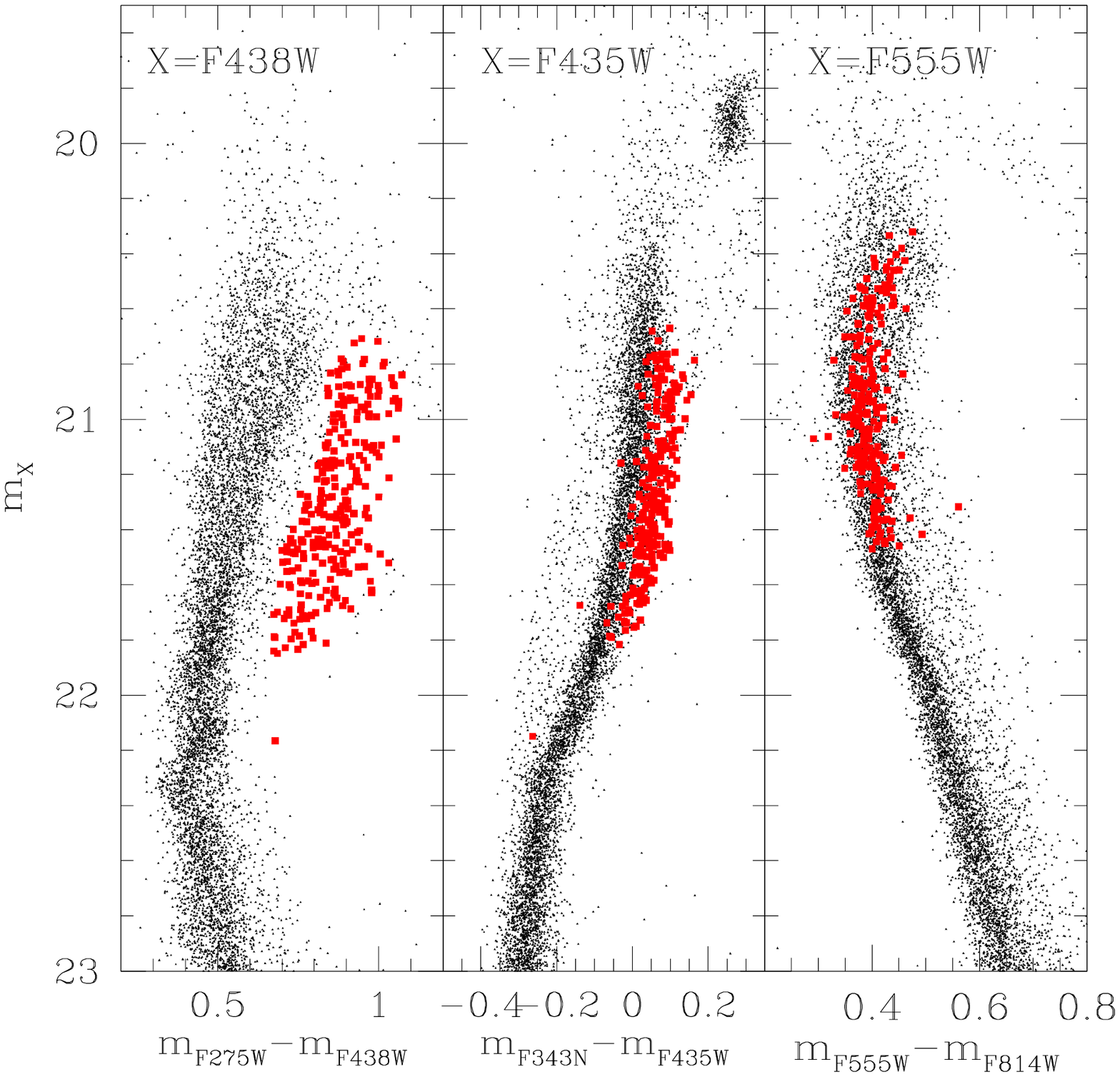}
\vskip -60pt
\caption{The sample of UV-dim stars highlighted by \citet{milone2022} in the UV CMD of the cluster NGC\,1783 is shown in the left panel. In the central and right panel the same stars are located in CMDs at increasing wavelength bands.}
\label{figmilone}
\end{figure}
 
The current literature on the CMD of young MC clusters is still debating the role of age spread and rotation, generally finding that variation in both parameters is necessary \citep[e.g.][and references therein]{gossage2019}. 
A breakthrough may come out from the new analysis of the stars in NGC\,1846 (age $\sim 1.5 \times 10^9$\,yr) by \cite{kamann2020}. By means of the MUSE integral-field spectroscopy, they measured the projected
rotational velocities of $\sim$1400 stars across the eMSTO and the upper MS, finding that the stars with larger $v \sin$\, i are clearly redder than those slowly rotating. Note that this MUSE sample is much larger than the previous spectroscopic data on the younger clusters with split MS quoted above, so in this case the possible role of braking is negligible, because the braked stars would anyway constitute only a few percent of the whole sample.
The \teff\ location of the most rapidly rotating MIST isochrones \citep{gossage2019} is redder than that of the isochrones with moderate or slow rotation rate, so the results of \cite{kamann2020} favour the `rotation' interpretation. Nevertheless, only structures rotating at 90\% of the critical rotation rate ---the limit at which the centrifugal force
equals the gravity of the star--- are sensibly redder than the slow rotating isochrones\footnote{This is expecially true for the range in rotation rates necessary to deal with the eMSTO when using the Geneva isochrones, while the effect in the MIST isohrones is already relevant at $\omega/\omega_c=0.6$, as shown in \cite{gossage2018}. Other solution may be available depending on the model assumed for the slower stars \citep[e.g.][]{wang2022}}. 

In a recent study, \cite{milone2022} examined the HST data for 113 clusters in the LMC and SMC and found a new feature coming out of the analysis of data including the UV band m$_{\rm F275W}$. In the plane m$_{\rm F438W}$ versus m$_{\rm F275W}$--m$_{\rm F438W}$, a non negligible number of stars is located sparsely beyond the fan, in a `red cloud' (Fig.\,\ref{figmilone}). We name this group `UV dim' stars. Notably, when we look at optical colors, e.g. at the CMD m$_{\rm F555W}$\ versus m$_{\rm F555W}$--m$_{\rm F814W}$, the UV-dim stars fall among the average colors of the eMSTO, indicating that their UV location is not related to the intrinsic  \teff. \cite{milone2022} also exclude that these stars represent the cooler tail of the distribution of rapidly rotating stars affected by limb gravity darkening and due to the projection effect. 

In this work we study the role of dust in shaping the eMSTO and show that the most likely reason for the observed spread and distribution of stars is the presence of  dust–like extinction.

 The outline of the paper is the following: In Sect.\,\ref{obsCMD} we put together a first exploration to understand the  eMSTO of NGC\,1783. First we go through the results of \cite{milone2022} and, by examining the CMD location of differently selected samples of eMSTO stars, we conclude that the wavelength dependent extinction may be due to circumstellar dust (Sect.\,\ref{CMDsamples}). 
We evaluate the  dust extinction in the spectra of stars having masses and \teff\ in the range of the TO stars in NGC\,1783  (Sect.\ref{dustmodels})  and use the results to model the colors of UV-dim stars. 
In Sect.\,\ref{models} we compare the CMD with different isochrone sets and show the limits of the isochrone fitting and of the interpretations. 
We then resort to semiempirical use of the data to make exploratory simulations of the dust effect (Sect.\ref{cmds}).
The location in the different CMDs reproduces well the data, providing a new interpretation for the extended MSTO, for puzzling features as the narrowness of the subgiant branch, and, mainly, for the correlation between rotation velocity and location in the eMSTO.  
In Sect.\,\ref{sim} we concentrate on the zig-zag feature present in the UV CMDs along the upper MS of several clusters, and in particular of NGC\,1783, and provide a working hypothesis to explain it.
We also perform simulations to begin to quantify the correlation between the zig-zag and the dust. Finally we discuss the results in Sect.\,\ref{discussion} and conclude that the circumstellar dust is a fundamental ingredient to model the eMSTO in clusters at intermediate ages as NGC\,1783 (1--1.8\,Gyr) and probably also in the younger clusters rich in Be stars. 

\section{Preliminary analysis and simple models for the turnoff fan of NGC\,1783}
\label{obsCMD}
\subsection{Selection of different samples of observed eMSTO stars}
\label{CMDsamples}
\begin{figure}
\centering
\vskip -50pt
\includegraphics[width=8.7cm]{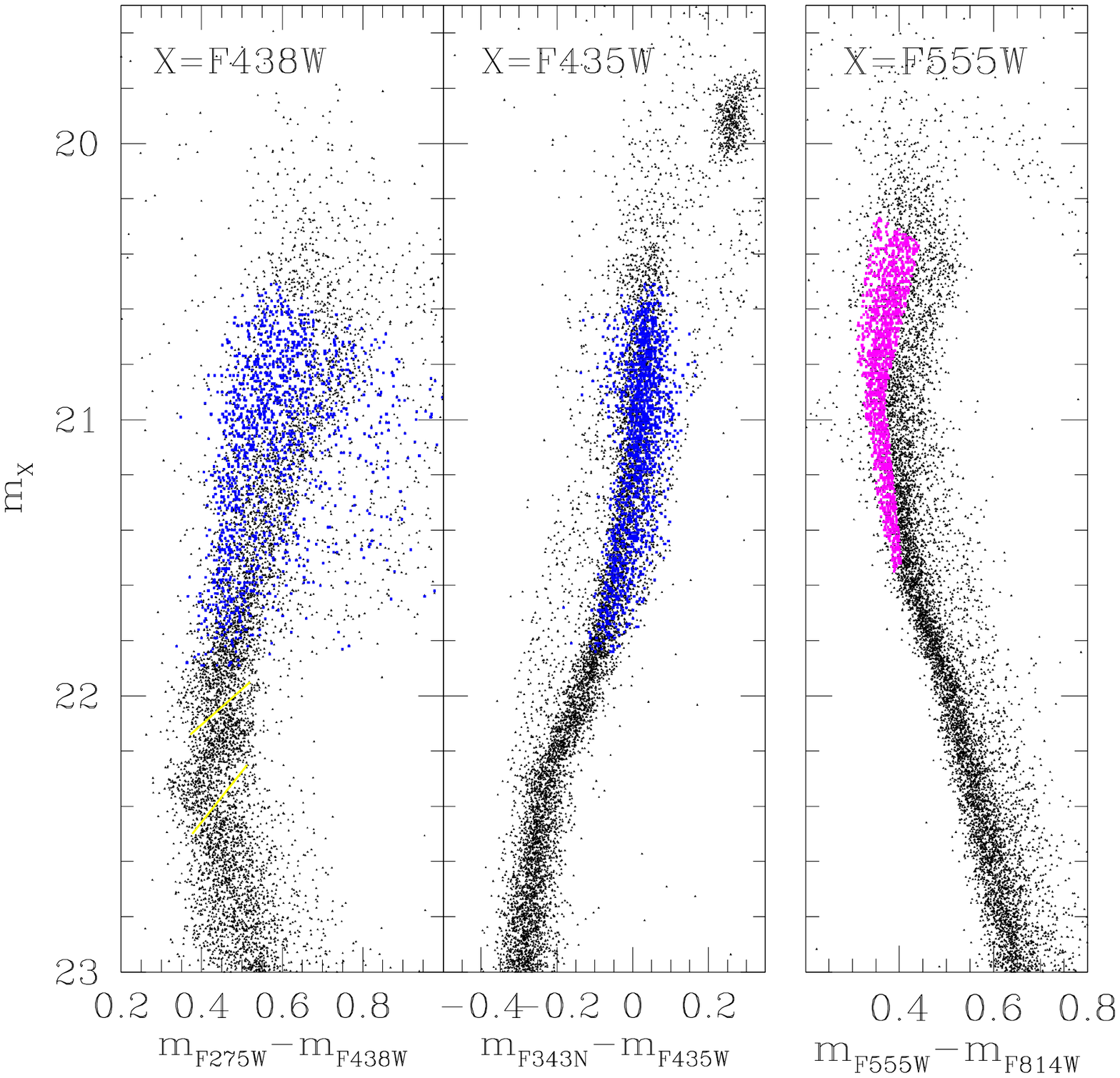}
\vskip -70pt
\caption{CMDs of NGC1783 from HST photometry. Here and in Fig.\,\ref{fig3}, \ref{fig6firstsim}, \ref{fig7sim} the two yellow diagonal lines in the left panel mark the two gaps visible in the stellar distribution only in this CMD. Magenta points mark a sample of eMSTO stars selected in the optical CMD (right panel). The location of these stars in the two other CMDs is marked by the blue points in the left and central panel. In the m$_{\rm F438W}$\ versus m$_{\rm F275W}$--m$_{\rm F438W}$\  the sample occupies the whole fan, and some stars are found to be UV-dim, out of the fan.}
\label{fig2}
\end{figure}

\begin{figure}
\centering
\vskip -50pt
\includegraphics[width=8.7cm]{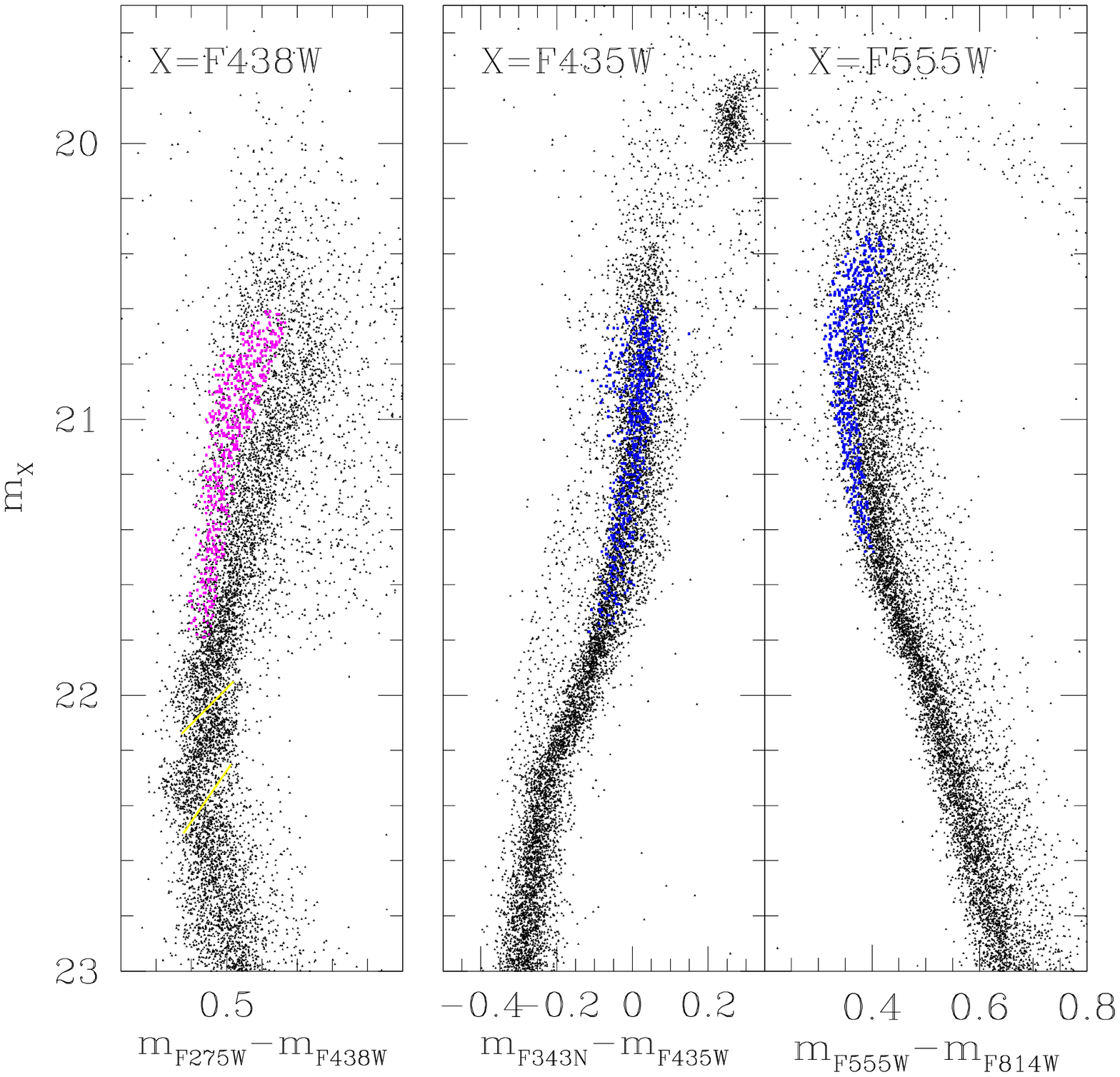}
\vskip -70pt
\caption{Reproduction of the CMDs of Fig.\ref{fig2}. The two yellow lines in the left panel mark the two gaps visible in the stellar distribution only in this CMD. Here, a sample (magenta dots) is selected from the  m$_{\rm F438W}$\ versus m$_{\rm F275W}$--m$_{\rm F438W}$\ CMD (left panel), and we see that they occupy similar locations, relative to the eMSTO, in the two other CMDs (blue points), indicating that this selection provides a sample not affected by UV absorption.}
\label{fig3}
\end{figure}

\cite{milone2022} discovered the cloud of stars redder than the eMSTO in the UV CMD and found out that these stars were not out of the eMSTO in the CMDs of filters at longer wavelengths (Fig.\,\ref{figmilone}). We begin our analysis by examining other samples of data: as the optical data are generally the most secure to derive the \teff\ and luminosity of stars, we select stars all lying in a strip on the left side of the fan in the CMD m$_{\rm F555W}$\ versus m$_{\rm F555W}$--m$_{\rm F814W}$\ (right panel of Fig.\,\ref{fig2})  and look at their location in the other CMDs: we see that these stars spread a bit out of the m$_{\rm F435W}$\ versus m$_{\rm F343N}$--m$_{\rm F435W}$\ diagram (central panel), but they are spread through the whole fan, {\it including locations among the UV-dim stars},  in the UV CMD m$_{\rm F438W}$\ versus m$_{\rm F275W}$--m$_{\rm F438W}$\ (left panel). This result shows that the eMSTO and UVdim stars are affected by a wavelength dependent absorption, so that the locations of stars in the optical–near IR CMD are not simply due
due to their \teff and luminosity.\\
Viceversa, if we select the stars on the left side of the UV CMD, these remain all in similar locations in the two other diagrams (Fig.\,\ref{fig3}). We tentatively conclude that the stars on the left side of the UV CMD represent stars whose colors are not affected by wavelength dependent absorptions.  We will discuss later on simulations starting from theoretical isochrones and, as first, we take a stellar sample like that isolated in Fig.\,\ref{fig3} as representative of the intrinsic CMD evolution of `standard' stars in the cluster.
\begin{figure}
\centering
\vskip -40pt
\includegraphics[width=7.5cm]{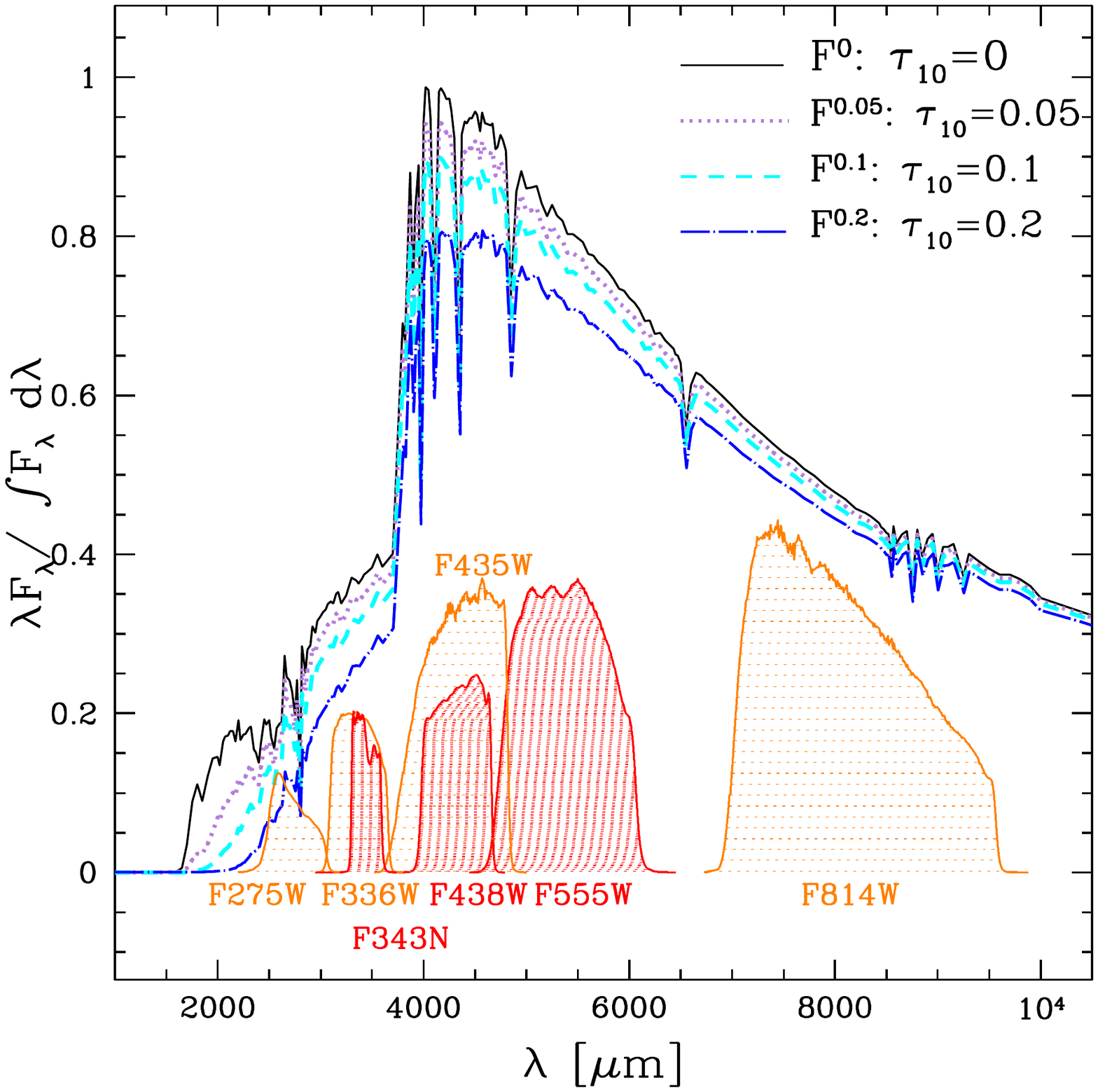}
\vskip -80pt
\includegraphics[width=7.5cm]{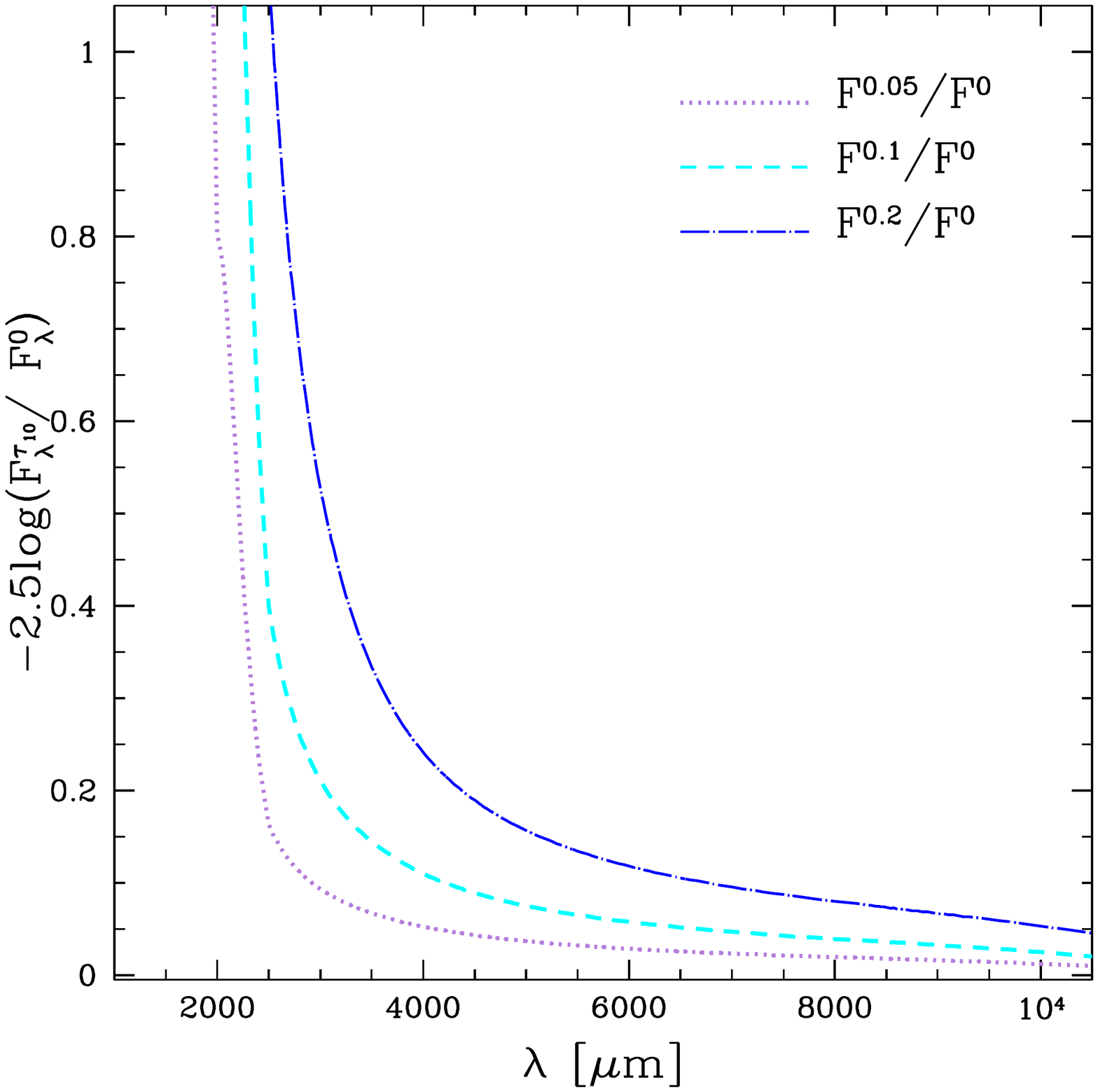}
\vskip -80pt
\includegraphics[width=7.5cm]{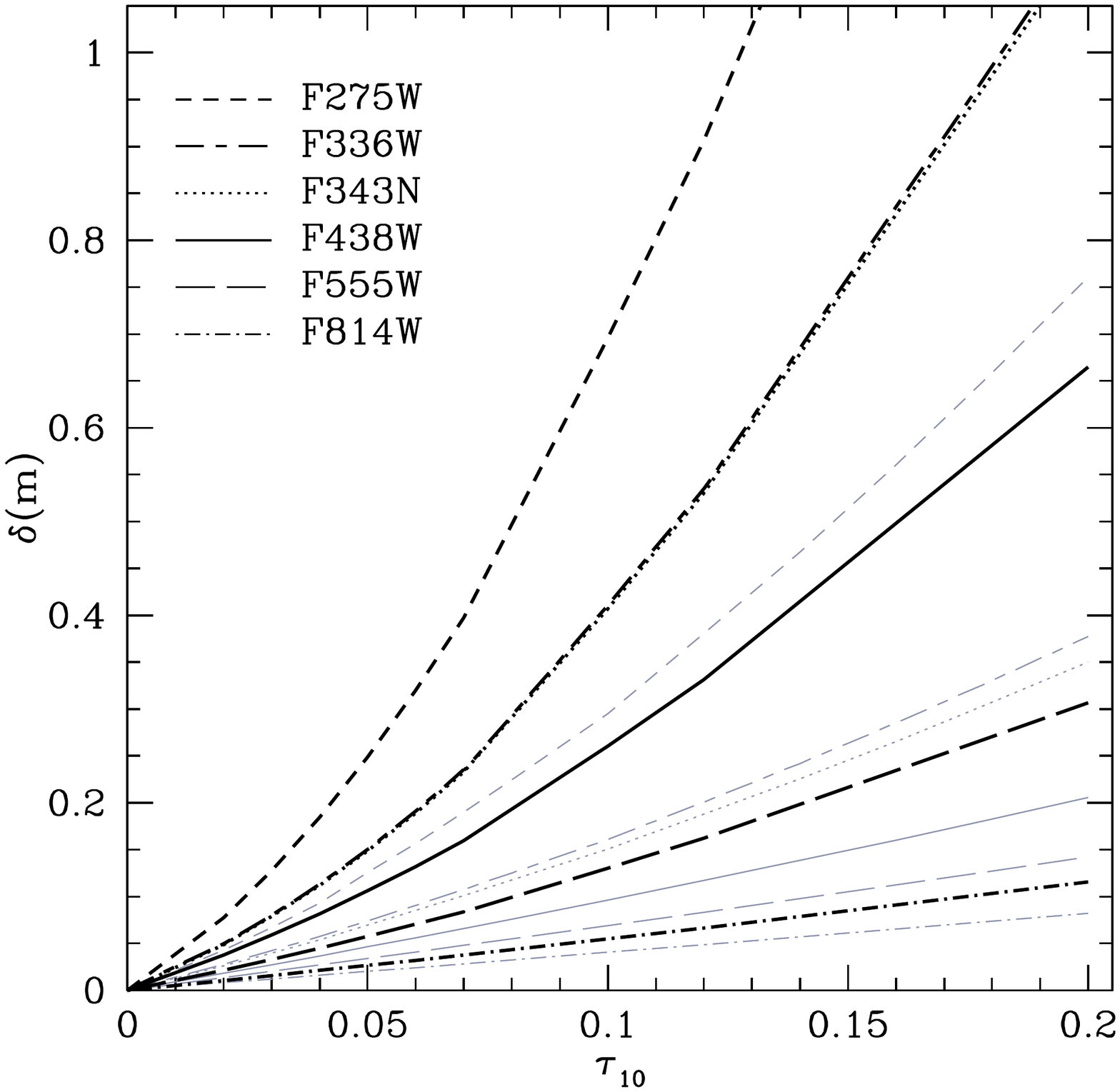}
\vskip -60pt
\caption{Top panel: The theoretical spectra of a star with \teff$=7500$K and $log(g)=4$, considering different level of dust absorption, assuming several optical depth values: $\tau_{10}=0$ (black solid line), 0.05 (violet dotted line), 0.1 (cyan dashed line) and 0.2 (blue dotted-dashed line). WFC3/UVIS transmission curves for the F275W, F336W, F343N and F438W filters and ACS/WFC transmission curves for the F435W, F555W and F814W filters are reported in orange and red. Central panel: the flux ratio between the theoretical spectra with optical depth $\tau_{10}>0$ and the spectrum with $\tau_{10}=0$.
Bottom panel: the absorption $\delta \rm m$ in each HST band as a function of the optical depth at 10 $\mu$m, $\tau_{10}$. Black lines refer to a grain size a=0.1$\mu$m, while gray lines are for smaller grains with a=0.05$\mu$m}
\label{fig4}
\end{figure}

\subsection{Role of dust}
\label{dustmodels}

From the observations of the young MC clusters, we know that large subsamples of their stars spin very rapidly. Among the younger clusters, where the MS and/or TO stars are still of B spectral type, many B--type stars with H$\alpha$\ emission (Be stars)  are photometrically detected \citep[e.g.][]{keller1999, bastian2017,correnti2017, milone2018, kamann2023}, especially among the upper MS stars. The Be-phenomenon is indeed linked with one or more mass-ejection processes, and observationally acts on top of a rotation rate of about 75\% of the critical rotation rate \citep{rivinius2013}. The rapid rotation is responsible for the formation of an excretion disk of gas in keplerian rotation, produced from material ejected from the central star. Spectroscopic confirmation of the H$\alpha$\ emission is found in NGC\,1866  \citep{dupree2017} together with a direct measurement of  rotational velocities in MS stars. Models have been built by \cite{georgy2013}. In a work devoted to explore the Be phenomenon, \cite{granada2013}  show that stars initially rotating close to the critical rate may reach it again during the MS evolution and lose mass. They found this effect also in the evolution of the smallest mass in their database, 1.7\Msun. 
As the stars evolving in NGC\,1783 have masses $\sim 1.5$\msun, we can make the hypothesis that they have lost matter in an excretion disk, and that this matter has condensed in typical silicate grains in the outer parts of the disk.  As dust is possibly the main wavelength dependent light absorber,  we begin modelling the effect of dust on the spectra of $\sim1.5$\Msun, having T$_{\rm eff}$ and gravity, g, in the range of the TO stars in NGC\,1783. We used the radiative transfer code DUSTY \citep{nenkova99} to simulate the reprocessing of the spectrum due to the presence of dust in the surrounding of the star. We started with a spectrum with T$_{\rm eff} =7500$K and $\rm log(\rm g)=4$ selected from the New Grids of ATLAS9 Model Atmospheres \citep{castelli2003}. We assumed the presence of silicates grains, the most stable dust species formed in oxygen-rich environments \citep{gail1999}, with a typical size of 0.05--0.1$\mu$m. We run DUSTY spanning a wide range of optical depths at 10 $\mu \rm m$, $\tau_{10}$, to consider different levels of dust absorption.

The typical absorption in the spectrum is shown in the top panel of Fig.\,\ref{fig4}.  From this and similar spectra computed for a variety of physical conditions in the range of the eMSTO stars of NGC\,1783, we derive typical absorptions in the HST bands as a function of $\tau_{10}$. These $\delta \rm (m)$\ for the relevant HST bands are plotted for two sizes of the grains (0.05 and 0.1$\mu$m) in the bottom panel of Fig.\,\ref{fig4}. Absorptions for the smaller size of grains (0.05$\mu$m) are similar to those that the larger size (0.1$\mu$m) achieves at a double $\tau_{10}$.  We adopt the grain size 0.1$\mu$\ for the first explorations (Sect.\,\ref{cmds}), and 0.05$\mu$m\ for the simulations in Sect\,\ref{simul}.  

\subsection{Theoretical models and their failures }
\label{models}
Attempts have been made to match the CMDs by a combination of variation in ages and rotation \citep[lately][]{gossage2019}, but the results remain ambiguous. The Geneva database of rotating models has been widely used to simulate the CMD of younger clusters, in particular those showing a split MS, for which the coexistence of a rapidly rotating and a slow rotating population seemed to be a plausible option \citep{dantona2015}. Nevertheless, this interpretation required that stars rotate very rapidly (close to break–up rotation) to fit the red MS and to explain the eMSTO extension as the result of projection effects of highly flattened stars. This requirement is at odds with a recent observational study showing that rotation velocities are slower than required, even in much younger clusters \citep{kamann2023}. 
The Geneva rotating computations end at 1.7\Msun, just above the typical masses evolving in this cluster. 
It is also well known that the Geneva rotating models and the MIST models \citep{dotter2016, choi2016}  provide results which differ enough from each other to be taken at face value. This may depend in particular on the different treatment of angular momentum transport in the two cases (advective vs. diffusive). 
\begin{figure}
\centering
\vskip -20pt
\includegraphics[width=7.5cm]{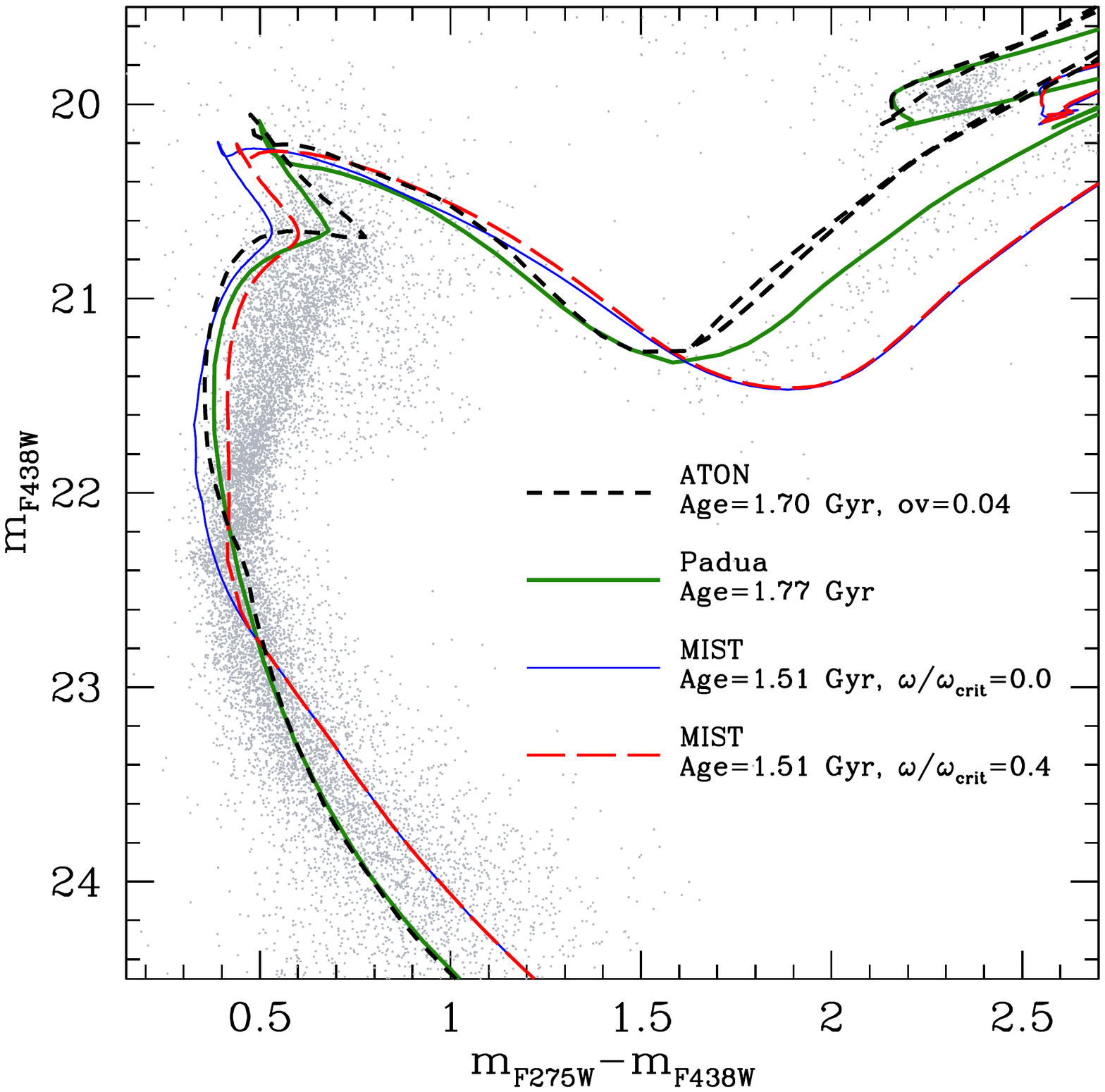}
\vskip -80pt
\includegraphics[width=7.5cm]{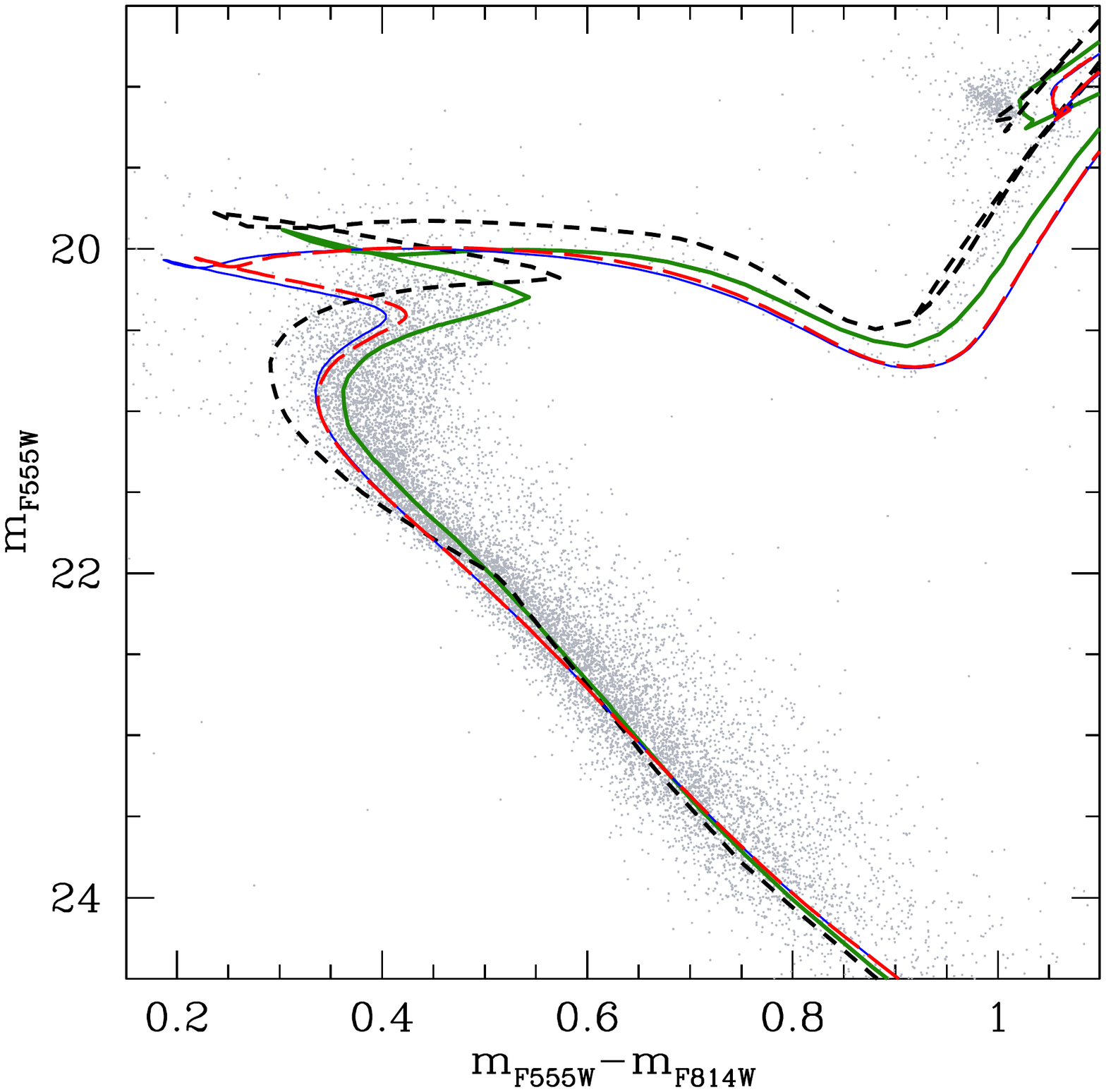}
\vskip -50pt
\caption{Top panel: m$_{\rm F438W}$\ versus m$_{\rm F275W}$--m$_{\rm F438W}$ diagram where the NGC 1783 data are shown as gray dots together with theoretical isochrones from different stellar evolution codes: ATON (black dashed line), PARSEC (green thick solid line), MIST with rotation (red long-dashed line) and without (blue thin solid line). The different parameters considered for each isochrone are reported in the legend. Bottom panel: the same as in the top panel reported in the m$_{\rm F555W}$\ versus m$_{\rm F555W}$--m$_{\rm F814W}$ diagram.}
\label{fig5cmds}
\end{figure}

Other available isochrone fitting include the `isochrone cloud' method by \cite{johnston2019}. The authors remark that, whatever the physics behind, the same mass can evolve with different convective core extension. For instance, this effect may be produced by different rotation rates and the consequent shear mixing at the border of the convective core, that can be modeled by different values of overshooting. If this is the case, the stars with smaller convective core evolve off the MS before the stars with larger cores, and, considering the cluster a coeval ensemble, the CMD will exhibit indeed a cloud of turnoffs which may resemble the eMSTO. Nevertheless NGC\,1783 (as well as many other MC clusters) shows a {\it very thin path of stars leading from the turnoff to the subgiant branch}, which resembles indeed a coeval cluster, but the isochrone cloud model would predict several different paths at different luminosity.

Comparing the CMDs of  NGC\,1783 with a few sets of available isochrones we recognize that single isochrones fail to match well the whole shape of the eMSTO. The top panel of Fig.\,\ref{fig5cmds} shows the UV based plane and the bottom panel shows the optical--near IR plane. In the comparison between NGC\,1783 data and the isochrones we assume the reddening E(B-V)=0.03 mag and the distance modulus (m-M)$_0$=18.51 \citep[see Table 2 in ][]{milone2022}. Three isochrones are shown for standard non--rotating models: the black dashed one is from our own database ATON \citep{tailo2020}, and will be adopted for the simulations in Sect\,\ref{simul}; the green thick solid one is  by the Padova database \citep{Marigo2017}, and the blue thin solid one is a MIST isochrone (slightly older than the other two, but having a similar turnoff luminosity). In addition, we plot in red long-dashed line the MIST isochrone for $\omega/\omega_{\rm crit}$=0.4\footnote{The MIST isochrones are very similar in the theoretical plane and in the optical colors, but they differ in the UV CMD. The MIST isochrones for  $\omega/\omega_{\rm crit}$=0.9 ---not displayed here--- are placed at redder colors \citep{gossage2019}.}. Although the turnoff shape is not well reproduced, both the Padova and ATON isochrone fit reasonably the luminosity and colors of the clump stars. The main difference between the two isochrones is the change of slope in the CMD when the stellar envelope becomes convective: the ATON models have a sharper transition from radiative to convective envelopes.
as they adopt the Full Spectrum of Turbulence (FST) description of convection \citep{cm1991, cgm1996}. 

This imperfect match between isochrones and data indicates that many factors contribute to the resulting eMSTO.
Consequently, we decided not to use the models to give a first look at the consequences of dust extinction on the observables (Sect.\ref{cmds}). The models will be needed to explore the motivation of the gap/peaks in the stellar distribution below the eMSTO (Sect.\,\ref{simul}).

\begin{figure}
\centering
\vskip -40pt
\includegraphics[width=8.5cm]{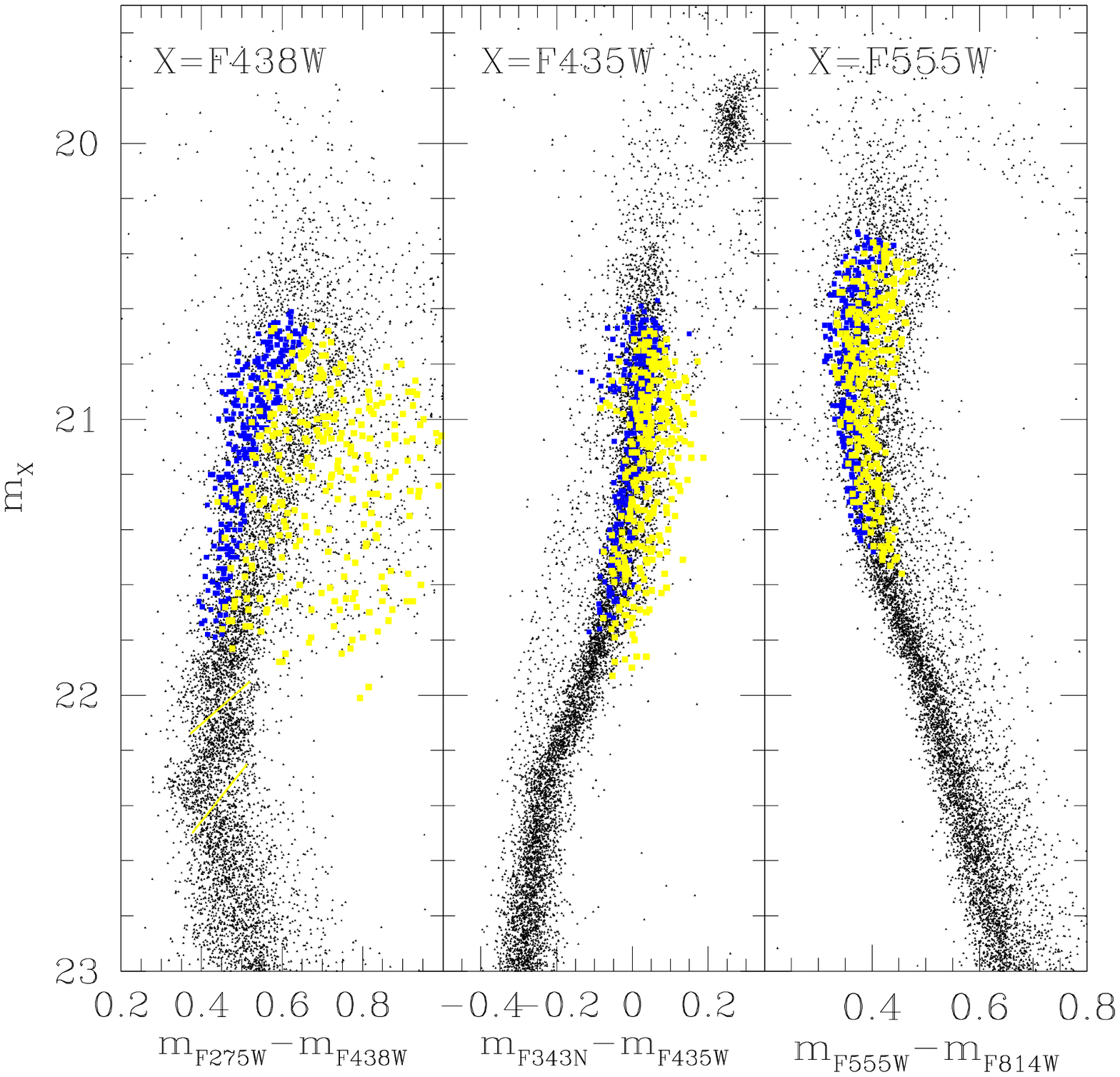}
\vskip -70pt
\caption{We show the effect of a random choice of dust absorption for  $0 \leq    \tau_{10} \leq 0.1$, reproducing the location of the UV-dim sample, but showing that also the fan locations may be affected by dust absorption. The starting sample is represented by the blue dots, selected in Fig.\,\ref{fig3}. The two yellow lines in the left panel mark the two gaps visible in the stellar distribution only in this CMD.}
\label{fig6firstsim}
\end{figure}

\section{Simple models for the eMSTO}
\label{cmds}
In this Section we outline the step-by-step path which leads to understand the basic role of dust for the eMSTO. We select in the UV CMD a group of stars which are in a `no dust' location, at the left side of the turnoff fan (blue squares in the left panel of Fig.\,\ref{fig6firstsim}) and look at their location under different assumptions.

\subsection{How to locate the UV-dim stars}
A first simple attempt to study the effect of dust is done by applying the absorption due to the presence of a dust disk to the sample located in the left side of the UV CMD, choosing random optical depth $0 \leq \tau_{10} \leq 0.1$\ (for grain size 0.05$\mu$). The results are shown as yellow dots in the right panels of Fig.\,\ref{fig6firstsim} for the three CMDs. We see that the stars with the highest $\tau_{10}$\ show up as UV-dim. These same stars can still be distinguished at the border of the eMSTO in the m$_{\rm F435W}$\ versus m$_{\rm F343N}$--m$_{\rm F435W}$\ diagram, but are all contained within the eMSTO in the  m$_{\rm F555W}$\ versus m$_{\rm F555W}$--m$_{\rm F814W}$\ CMD. Thus the dust extinction effect can reproduce the location the UV-dim sample  (Fig.\,\ref{figmilone}) in all CMDs. \\
While the UV-dim sample is reproduced with the highest  $\tau_{10} $\ values, lower extinctions obviously produce objects within the fan. Nevertheless, even with the highest $\tau_{10}$'s many stars must be less absorbed, because the dust around the star is likely distributed in a ring or a disk, so the absorption will depend on the viewing angle, that is on the inclination i of the rotation axis (perpendicular to the disk plane) with respect to the observer: for i=90$^\circ$, the extinction is maximum, while for  i=0$^\circ$, the star is viewed pole-on and is not obscured. A simple first estimate of the angle of view effect is obtained by multiplying the extinction by the sin\,i of a randomly chosen angle i. The consequence is non trivial: when we attempt to describe the UV-dim sample by dust absorption, at the same time  {\it many of the stars having dust disks will be located in the eMSTO, because they are seen at a low inclination angle and have a lower extinction.} Is it possible that the {\it whole} fan morphology is due to dust obscuration? 

\begin{figure}
\centering
\vskip -50pt
\includegraphics[width=8.5cm]{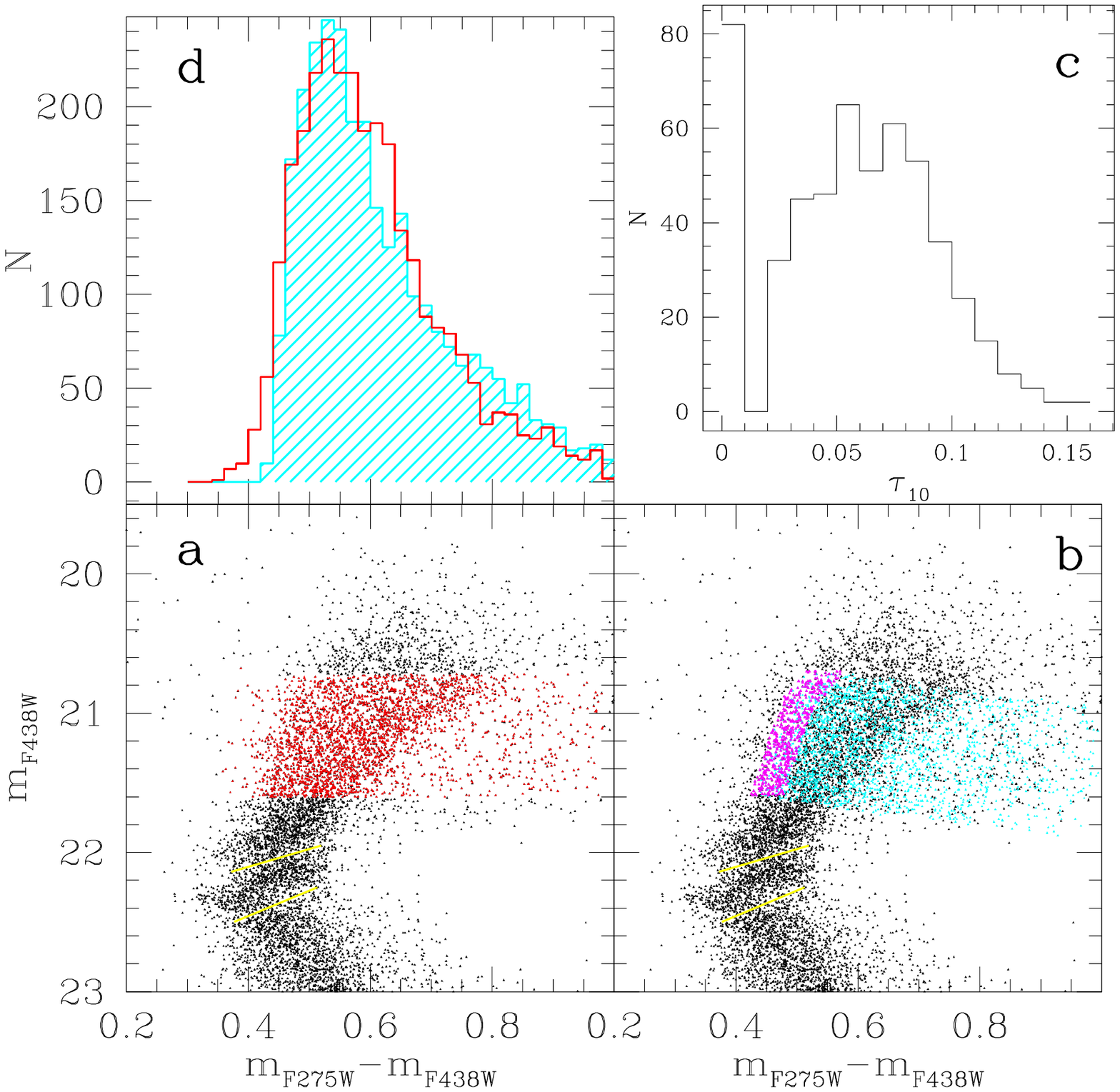}
\vskip -70pt
\caption{ Panel a shows the observational UV CMD, and highlights in red the stars in the range 20.7$\leq$m$_{F438W}\leq$21.7. Panel b shows the same CMD and superimposed the sample chosen as starting points for the simulation (magenta dots) and the result of the simulation (cyan). The two yellow lines in panels a and b mark the two gaps in the stellar distribution. Panel c shows the optical depth distribution adopted in the simulation: a gaussian with mean $\tau_{10}= 0.055$\ and $\sigma$=0.035. For $\tau_{10}<0.02$\ the optical depth is truncated to 0. In panel d we compare the observed and simulated distribution.}
\label{fig7sim}
\end{figure}
\begin{figure}
\centering
\vskip -50pt
\includegraphics[width=8.25cm]{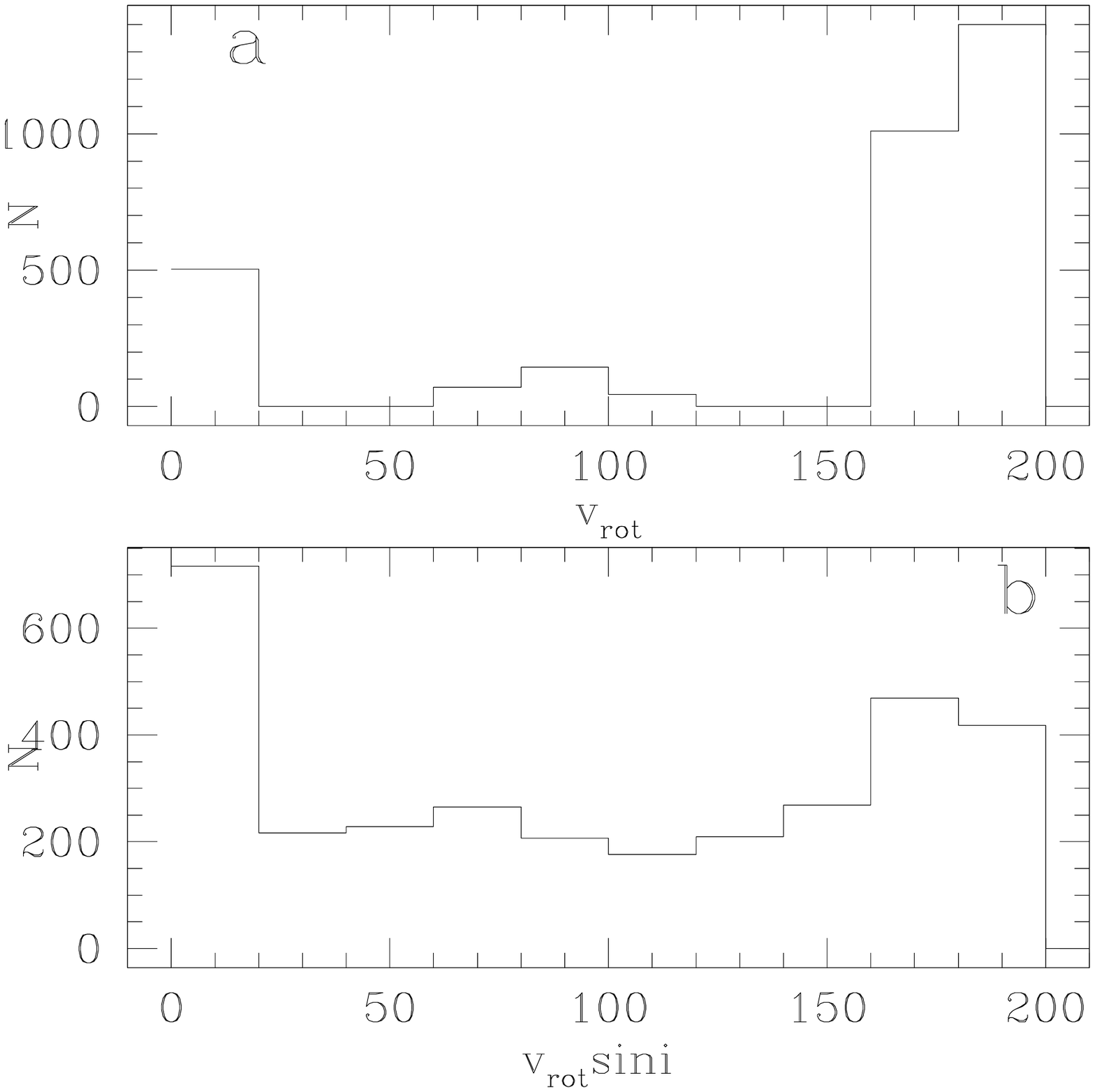}
\vskip -70pt
\caption{Distribution of rotation rates adopted in the simulation (panel a): we have many stars with $v_{\rm rot}$=0, corresponding to those with no dust, some intermediate velocities and then two groups at $v_{\rm rot}$=180\,Km/s and 200\,Km/s. Panel b shows the observational $v_{\rm rot}$sin\,i, where the inclination is the same which enters in the computation of the dust absorption.}
\label{fig10rot}
\end{figure}

\begin{figure*}
\centering
\vskip -30pt
\includegraphics[width=8.5cm]{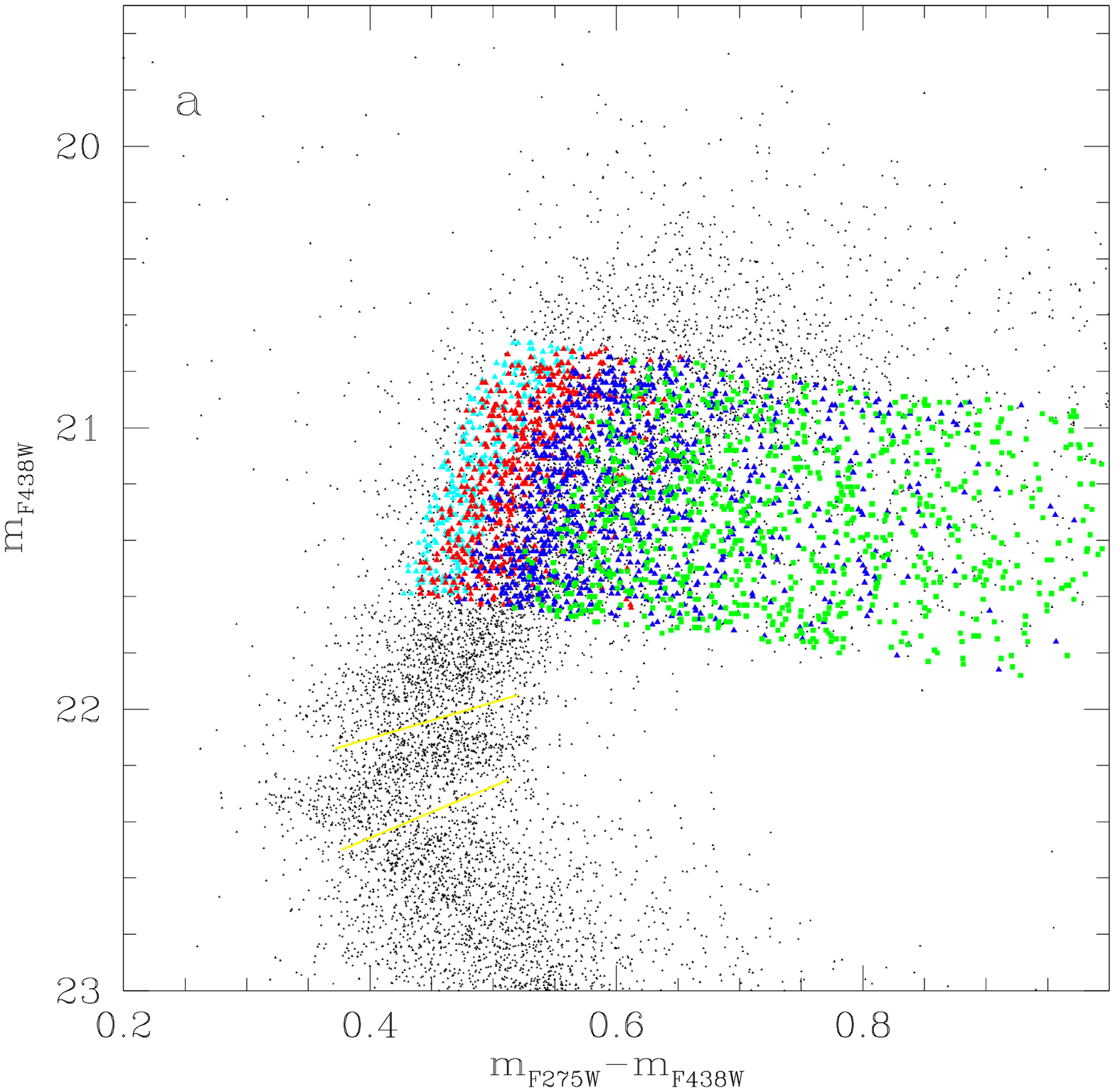}
\includegraphics[width=8.5cm]{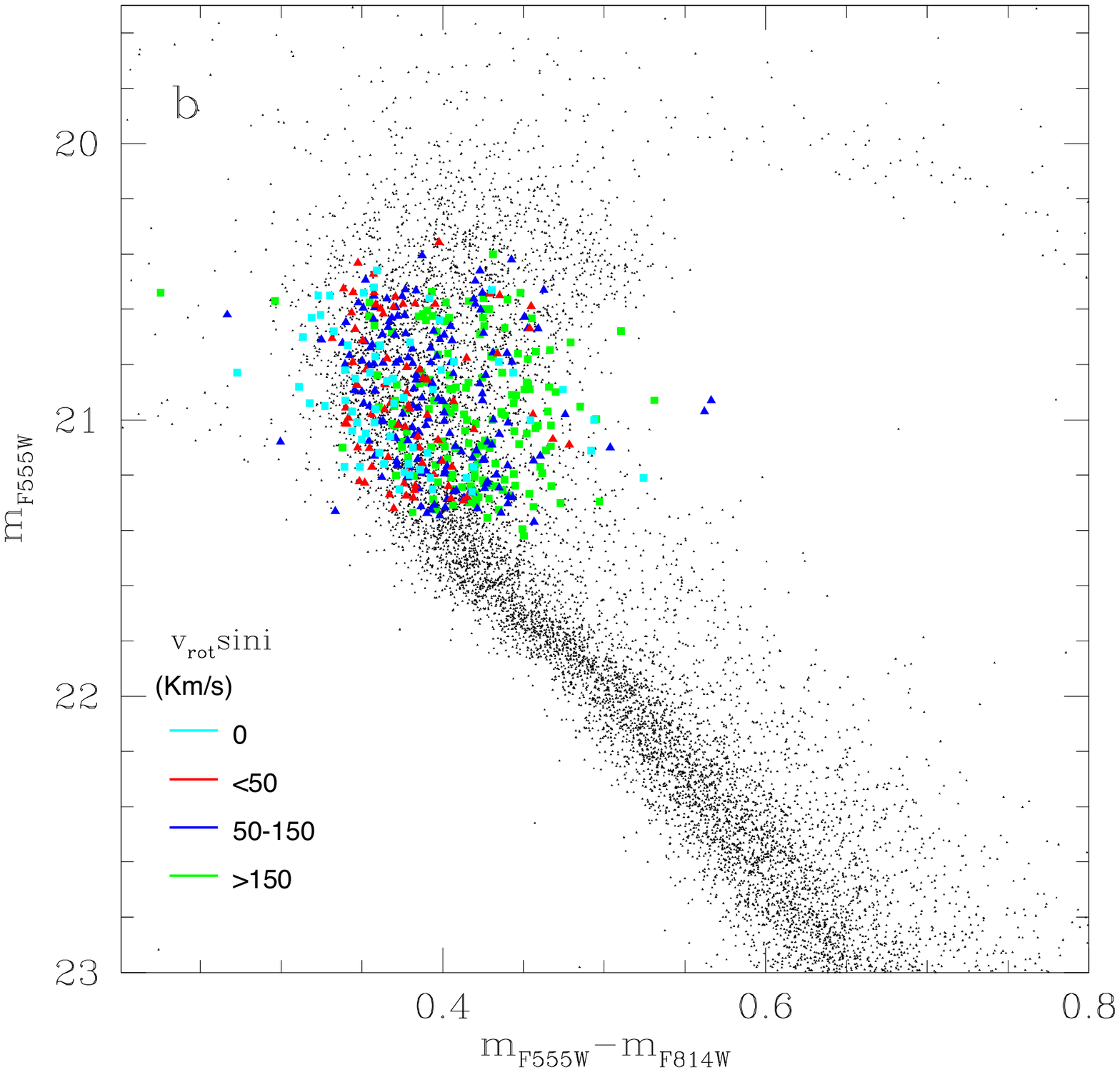}
\vskip -70pt
\caption{The simulation is shown in the UV CMD (panel a) and in the visual-near IR CMD (panel b), highlighting the distribution of stars with different rotation observable rates. The highest velocity group is on average located at redder colors than the others, as we have attributed larger rotation rates to the stars with larger  $\tau_{10}$. In particular, the UV-dim sample should contain mostly high rotation stars. }
\label{fig11simrot}
\end{figure*}

\subsection{A more detailed simple model }
\label{simplemodel}
We implement a less trivial distribution of $\tau_{10}$. For each magnitude band m$_{\rm intrinsic}$, the magnitude of the star, when obscured by dust, m$_{\rm abs}$, will be
\begin{equation}
\rm m_{abs} = m_{intrinsic}+\tau_{10} \frac {\delta m_{band}}{ \delta \tau_{10}}  \sin \rm i
\label{eq1}
\end{equation}
where we adopt a truncated gaussian distribution for $\tau_{10} $, and a random value between 0 and $\pi/2$\ for the inclination. For each band, $\rm \delta m_{band} / \delta \tau_{10}$ is taken from Fig.\,\ref{fig4}.
We truncate the distribution of dust below a small value of $\tau_{10} $, assuming that there is a (more or less relevant) percentage of stars not affected by dust.
 
The distribution of dusty stars will resemble the observational fan: the left side will be populated by the stars with $\tau_{10} $=0, but  there will be also a right side border, located at the maximum of the $\tau$\ distribution and inclination close to i=90$^\circ$.  Intermediate locations are populated both by stars with smaller values of $\tau$\  and those with lower inclinations. The UV-dim stars are reproduced by the tail of the gaussian $\tau$\ distribution: as these represent a small percentage of the fan stars, the $\sigma$\  of the distribution must be quite small\footnote{It is also possible that the $\tau_{10}$\ distribution is not gaussian but more skewed.}.

In Fig.\,\ref{fig7sim} we plot the results of a simulation in the UV CMD. Panel a shows the observational diagram and highlights in red the eMSTO and UV-dim stars having   20.7$\leq \rm m_{F438W} \leq$21.7. Panel b shows the simulation (cyan dots) obtained by starting from the sample of stars highlighted in magenta, assuming a gaussian $\tau_{10}$\ distribution with mean value $\tau_{10}$=0.055 and $\sigma$=0.035, truncated below $\tau_{10}$=0.02  (panel c). The comparison between the histogram of observations and simulation (panel d) shows that the assumptions of the model are reasonable and that {\it the whole fan spread may be due to dust and not to different evolutionary paths}.  

This comparison is very crude, as it is done on the whole magnitude range of the fan, starting empirically from the location of the left boundary of the fan itself.  
To improve the fit we need to slightly increase the mean $\tau_{10}$\ value for increasing luminosity along the eMSTO (see Section\,\ref{simul}).

\subsection{Further results of the simple model: the stellar rotation rate and dust absorption}
We may guess purely semiempirically the behaviour  of another physical parameter,  the stellar rotation rate. 
It is plausible that stars rotating below a fixed velocity threshold do not lose matter at all, so they will not form a dust ring and will not be obscured. Above this threshold we resort to assume $\tau_{10} \propto v_{\rm rot}$. Obviously, the observable will only be the projected velocity, $v_{\rm rot}$\,sin\,i. We make the implicit but reasonable assumption that the inclination angle is the same for the disk ---and for the dust disk or ring--- and for the spin, as we expect that the mass loss takes place at the equator.
In Fig.\,\ref{fig10rot}  we show a possible assumption for the distribution of rotational velocity in the simple gaussian simulation (panel a), and the effect of projection on this distribution (panel b). While the qualitative behaviour is reasonable, the precise values of velocities are chosen only as an example, but the observational consequence holds for any distribution and is shown in Fig.\,\ref{fig11simrot}: {\it the higher velocity stars will be preferentially located at redder colors in the fan}, because the assumption that high initial rotation goes together with more mass loss (and more dust) is also amplified by the inclination effect: the stars having the fastest rotation rate and  seen with high inclination will be most affected by dust (redder) and also show the highest $v \sin$\,i.  This result may provide a plausible but alternative explanation for a result often found in the observations \citep{dupree2017, marino2018a, marino2018b, bastian2018}, but especially confirmed for the cluster NGC\,1846 \citep{gossage2019} and for the much younger cluster NGC\,1850 \citep{kamann2023}, that the stars with larger $v \sin$\, i are found preferentially in the red side of the eMSTO.  The cooler location of rotating isochrones may be a concomitant, but not necessarily the dominant, explanation for such a result.  

\begin{figure*}
\centering
\vskip -50pt
\includegraphics[width=13.5cm]{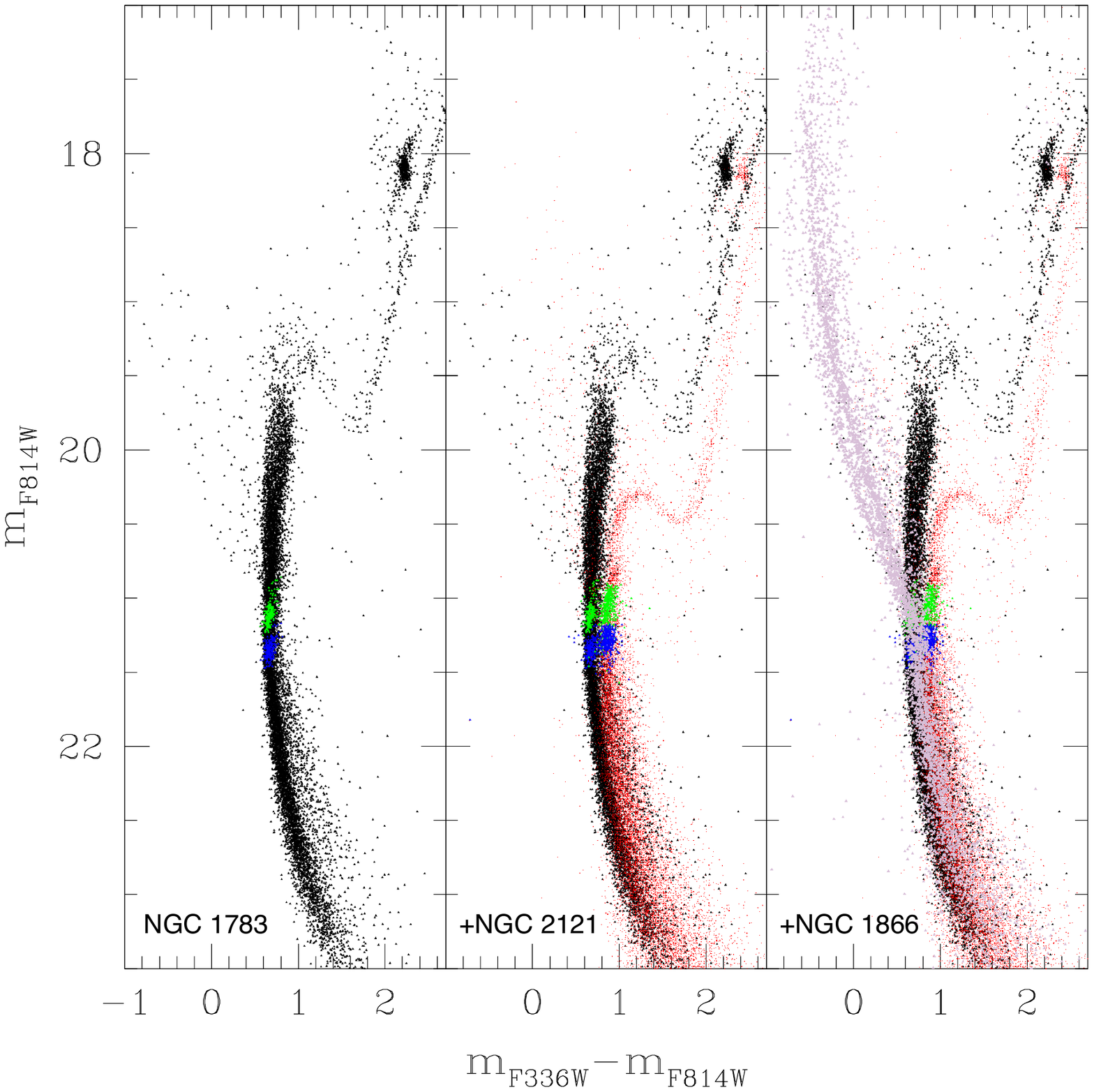}
\vskip -100pt
\caption{Comparison of the CMD of NGC\,1783 with the older cluster NGC\,2121 and the younger clusters NGC\,1866. The peaks in the stellar distribution of NGC\,1783 and NGC\,2121 are highlighted in green (upper peak) and blue (lower peak). The comparison with NGC1866 shows that they are located close to the region where the MS changes slope, when the stars begin to have convective envelopes. It is suggested that the upper peak is related to this transition for the most rapidly rotating stars, which populate the more luminous MS (as displayed by NGC\,1866 stars) and the lower peak is due to clustering of `braked' stars (see text). 
mostly high rotation stars. }
\label{fig12comppeaks}
\end{figure*}
\section{The zig-zag features in the MS stellar distribution}
\label{sim}
This possible explanation of the eMSTO and the UV-dim stars in NGC\,1783 leads us to extend the model and look at the peaks \& gaps seen in the upper MS of this cluster and of other clusters of intermediate (1.2-1.8\,Gyr) age \citep{milone2022}. The location of the gaps is highlighted in yellow in the left panel of  Fig.\,2: these features are strikingly prominent in the CMD in the UV colors, although in the optical CMDs the same stars are spread out on a wider range of magnitudes, diluting the small discontinuities. 

According to our model, the dust surrounding the stars in the eMSTO of NGC\,1783 is circumstellar, a consequence of mass loss occurring during the core hydrogen burning stage. This phenomenon has been discussed at long in the literature, in connection to the Be stars. 
It is a largely accepted idea that the Be stage occurs in very rapidly rotating stars, which are surrounded by a gaseous, geometrically thin excretion disk in near Keplerian rotation. As all stars down to the masses whose envelope in MS becomes convective may be fast rotators \citep[see, e.g., ][]{zorec2012}, the phenomenon extends to the stars populating the eMSTO of NGC\,1783. As a guideline, we consider the models by \cite{granada2013}, describing the mass loss phase leading to the formation of an excretion disk. The authors assume that intense mass loss occurs from the time when the angular velocity  $\omega$\ reaches 99\% of the critical rotation velocity, $\omega_{\rm max}=\omega/\omega_{\rm crit}=0.99$, and that the mass ejected mechanically from the model is governed by the quantity of angular momentum that needs to be removed from the surface to maintain the rotation rate below the limit set by this chosen maximum $\omega$. Thus mechanical mass loss goes on until the transport of angular momentum by meridional circulation and shears increase the rotation rate again up to $\omega_{\rm max}$. Episodes of mass loss will occur till the end of the core hydrogen burning. The total amount of mass lost by the lowest mass model (1.7\Msun) computed by \cite{granada2013} is $\sim 6 \times 10^{-3}$\msun\ for metallicity Z=0.006.
Although these models are indeed a useful guideline, we should be aware that the modalities of mass loss and the rotation limit above which it occurs are not necessarily limited to these specific choices. In particular, we know that the Be stars do not rotate as close to break-up as suggested, so the mass loss phase may begin well before $\omega_{\rm max}$=0.99, or it may be more intense and carry away more angular momentum than modeled here.  

\subsection{A working hypothesis for the zigzag}
We take now an observational approach to the scenario: the distribution of stars along the upper main sequence shows two regions alternating  over-density and under-density, in a zig-zag shape. This feature is also present in other clusters, as  described in \cite{milone2022}, and is mainly detected in the m$_{\rm F438W}$ versus m$_{\rm F275W}$--m$_{\rm F438W}$ CMD. 
In the left panel of Fig.\,\ref{fig12comppeaks} we show a standard m$_{\rm F814W}$ versus m$_{\rm F336W}$--m$_{\rm F814W}$  CMD of the cluster and highlight in green and blue the location of the two peaks in the stellar distribution\footnote{The stars in zig-zag feature have been marked in the UV CMD, no peak is seen here.}. In the central panel we superimpose the CMD of NGC\,2121, an older cluster also showing prominently the peaks (see Fig.\,4 in \cite{milone2022}).
), which now appear at a slightly lower \teff. A younger (initial) MS location is represented by adding the CMD of NGC\,1866 in the right panel. From the latter comparison we see that the peaks in NGC\,1783 are indeed close to the \teff\ where the MS changes slope both in the CMD and in the mass luminosity relation, due to rapid the transition between models having radiative and convective envelopes.  
At \teff$\sim$6800\,K a gap was recognized in the CMD of the Hyades \citep[and recently  confirmed for the Hyades in the Gaia data][] {reino2018}, and interpreted  by \cite{dantona2002bv} as a `\teff\ gap'. Another gap is known, at $\sim 7250$,  the classic B\"ohm-Vitense gap \citep{bvcanterna1974} attributed by Erika B\"ohm-Vitense to the onset of convection in the deep atmospheric layers of these stars. Here the MC cluster data may suggest a different interpretation.

We have shown that  the presence of the eMSTO and the UV-dim stars in NGC\,1783 require as a plausible explanation that  (a consistent fraction of) the upper MS stars have lost mass mechanically at the formation of an excretion disk due to rotation. But where the envelope is convective, any initial fast rotation will slow down, e.g. by magnetic braking, so that mechanical mass loss can not occur any longer. 
We suggest that either one or both peaks \& gaps in the star distribution be due to the transition from stars losing mass in an excretion disk to stars which continue their evolution without significant mass loss. In the  transition region from stars losing mass to stars preserving their initial mass, a peak in the stellar distribution will result.  The peak will be due to summing up of the mass loss events shown in the \cite{granada2013} models, but all occurring during the long MS lifetime of these lower mass stars ($\sim$1.3\msun), with the peak \& gap evolving together with the isochrone, as shown by the shifting of the structures passing from NGC\,1783 to NGC\,2121.

Now we ask whether we can estimate the total amount of mass loss which is compatible with the peaks. 
Simple stellar models show that a fast (on thermal timescale or shorter) mass loss starting at a precise point of the MS life shifts the star evolution simply to the luminosity and \teff\ of the model of the lower mass at the same stage of nuclear evolution. Therefore, if, say, a 1.5\Msun\ star loses 0.005\Msun\ when its core hydrogen is content X$_{\rm c}$=0.4, it will find itself at the location of the 1.495\Msun\ of about the same age and X$_{\rm c}$=0.4. This rough approximation may give an order of magnitude idea of the amount of mass loss needed to reproduce the two peaks/gaps. 
As we have seen, the available isochrones do not reproduce well the shape of the eMSTO left boundary, so it will be necessary to compare the results to a limited range of magnitudes.
A more detailed comparison involves the computation of tracks and isochrones for the evolution of rotating stars which reach the break up velocity, such as in \cite{granada2013}, but for the masses and chemistry appropriate to this cluster. 

\subsection{Simulations}
\label{simul}
The idea that the presence of zig-zag features is the result of the mass loss experienced by stars with a radiative envelope can be supported by the computation of a synthetic population. To this scope, we compute the synthetic population by randomly extracting stars over an initial mass function (IMF) and distributing them along the luminosity function obtained from the isocrones.    
 
In absence of proper rotational isochrones, and with the aim to consider carefully the dimmer gap as done in \cite{dantona2002bv}, we adopt the ATON database \citep{tailo2020}, from which we select the isochrone of 1.7\,Gyr for metal mass fraction Z=0.0065, [$\alpha$/Fe]=+0.2 helium mass fraction Y=0.26, shown dashed in black in Fig. \ref{fig5cmds}. Convection is treated according to the Full Spectrum of Turbulence (FST) model by \cite{cgm1996}, core overshooting is fixed to $\zeta$=0.04 in the diffusion approximation described in \cite{ventura2008}. The chosen core overshooting parameter is larger than assumed in standard computation ($\zeta$=0.02) to simulate the possible effect of extra mixing at the border of the convective core due to rapid rotation.  

\begin{figure}
\centering
\vskip -50pt
\includegraphics[width=8.5cm]{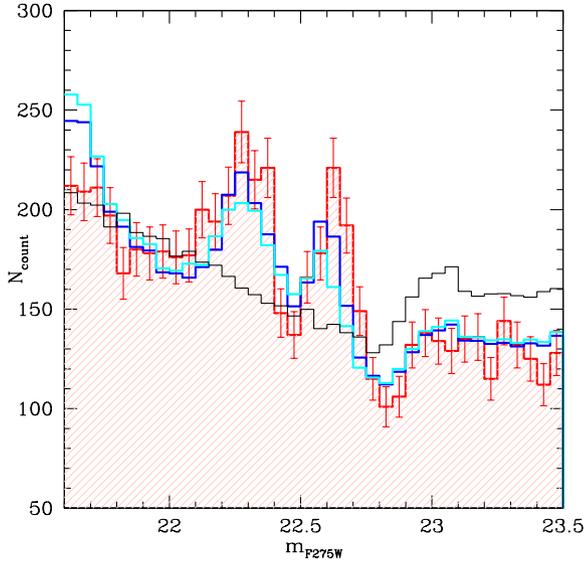}
\vskip -50pt
\caption{Comparison of the histogram of stellar counts as a function of
m$_{\rm F275W}$ between the data (red) and simulations without mass loss (thin black), with mass loss (blue) and
with reduced mass loss (cyan) (see Section \ref{simul}). The simulation is normalized to the number of stars counted in the magnitude range shown and within $0<$m$_{\rm F275W}$--m$_{\rm F438W}<0.9$.}
\label{figisto}
\end{figure}

We use as basis for the comparison the histogram distribution of the stars versus the m$_{\rm F275W}$\ magnitude, where the two peaks and gaps appear clearly separated.
The standard luminosity function in the m$_{F275W}$\ band is obtained by convolving the mass {\it vs. } m$_{F275W}$\ band relation from the isochrone with a mass function dN/dM$\propto M^{-1.1}$. The result is shown as a black histogram in Fig. \ref{figisto}, where normalization is made considering the total number of stars in the magnitude interval 21.5--23.5 and in the color interval $0<$m$_{\rm F275W}$--m$_{\rm F438W}<0.9$. The simulation does not show the two observed peaks, although it reasonably reproduces the dimmer gap in the data. This gap is indeed the signature of the change in mass--luminosity and mass--\teff\ relation when the envelope becomes convective. 

In order to account for the peaks ---although it is clear that using a single isochrone is a very primitive option--- we introduce two elements in the simulation:
\begin{enumerate}
\item We consider the effect of the rotation and therefore of the mass lost. Once obtained a distribution of initial masses for the synthetic population, we assume that the stars evolving above the fainter gap have lost mass due to their high velocity rotation during their lifetime. We subtract the total amount of mass, $\Delta (mass)$, lost by each star, to the initial mass and locate the final mass along the isochrone. We guess $\Delta(mass)$\ by adjusting the resulting peaks and gaps shown in the Fig. \ref{figisto} and observed in the CMDs. 
\item We consider the effect of the dust absorption. The magnitudes of the stars are computed applying the approach described in Sect.\,\ref{simplemodel} {\it to the stars that have lost mass and should be surrounded by a ring of dust}. $\delta \rm m_{\rm band}$ is computed as described in Eq.\,\ref{eq1} for all the stars evolving above the fainter gap up to the TO. Along our case isochrone, this includes stars with initial mass in the range $1.2$M$_{\odot}<$M$_{\rm in}<$1.6M$_{\odot}$. For the stars not belonging to this range of masses $\delta m_{\rm band}=0$.  $\tau_{10}$ is randomly assigned and distributed along a gaussian curve, whose mean value and $\sigma$ are chosen from the comparison with the distribution of the stars in different CMD combinations. The relation between $\tau_{10}$ and $\delta \rm m_{\rm band}$ is obtained from the synthetic spectra with absorption, as described in Sect.\,\ref{dustmodels} and shown in Fig.\,\ref{fig4}. 
\end{enumerate}
In Fig. \ref{figisto} the 
blue histogram shows a simulation including  $\Delta(mass)=0.008$M$_{\odot}$ for each star with mass $1.3\rm M_{\odot}\leq$M$_{in}\leq1.6$M${\odot}$, $\Delta(mass)=0.004$M$_{\odot}$ for the range 1.26--1.3\Msun, and $\Delta(mass)=0.006$M$_{\odot}$ for the range  $1.24\rm M_{\odot}\leq$M$_{in}\leq1.26$M${\odot}$. 
All these stars evolve in the range of magnitudes $21.5 <$m$_{\rm F275W}<23.5$, and the range 1.26--1.3\Msun\ is located in correspondence with the upper gap at m$_{\rm F275W}\sim22.4$mag. In the cyan histogram in Fig. \ref{figisto}  the mass loss is reduced by a factor ten (i.e. $\Delta(mass)=0.0008$M$_{\odot}$, $\Delta(mass)=0.0004$M$_{\odot}$ and $\Delta(mass)=0.0006$M$_{\odot}$).

As we know, the MS of younger clusters is split (as also seen in the plot of the CMD of NGC\,1866) and the more luminous MS should be the locus of stars in fast rotation, precisely those subject to form the excretion disks. Therefore, it is the upper peak--gap which we can describe as a result of mass loss. This peak is also due to the inclusion of dust absorption in the simulation, and in fact it is not so affected by a reduction of the mass loss by an entire factor 10. \\
As a further speculation, we suggest that a fraction of these sub-critically rotating stars are fully braked as result of the mass loss, an effect similar to what produces the apparently younger upper MS in several clusters \citep{dantona2017}. As a consequence, they accumulate in the MS location of non rotating stars, so explaining the presence of the dimmer peak.
If this were the case, we should find out that all the stars in the dimmer peak have low rotation velocities, while the stars in the upper peak should be a mixture of rapid rotating and slow rotating stars.  \\
It is also possible that the dimmer gap reflects the effect described by \cite{dantona2002bv}, and in fact it is also seen in the standard luminosity function (black histogram in Fig.\ref{figisto}).
Much more work will be needed to get a satisfactory whole picture of the process, possibly extending the analysis also to the stellar distribution in other clusters.

\subsection{Comparison between the simulated and the observed CMDs}
\begin{figure}
\centering
\includegraphics[width=8.5cm]{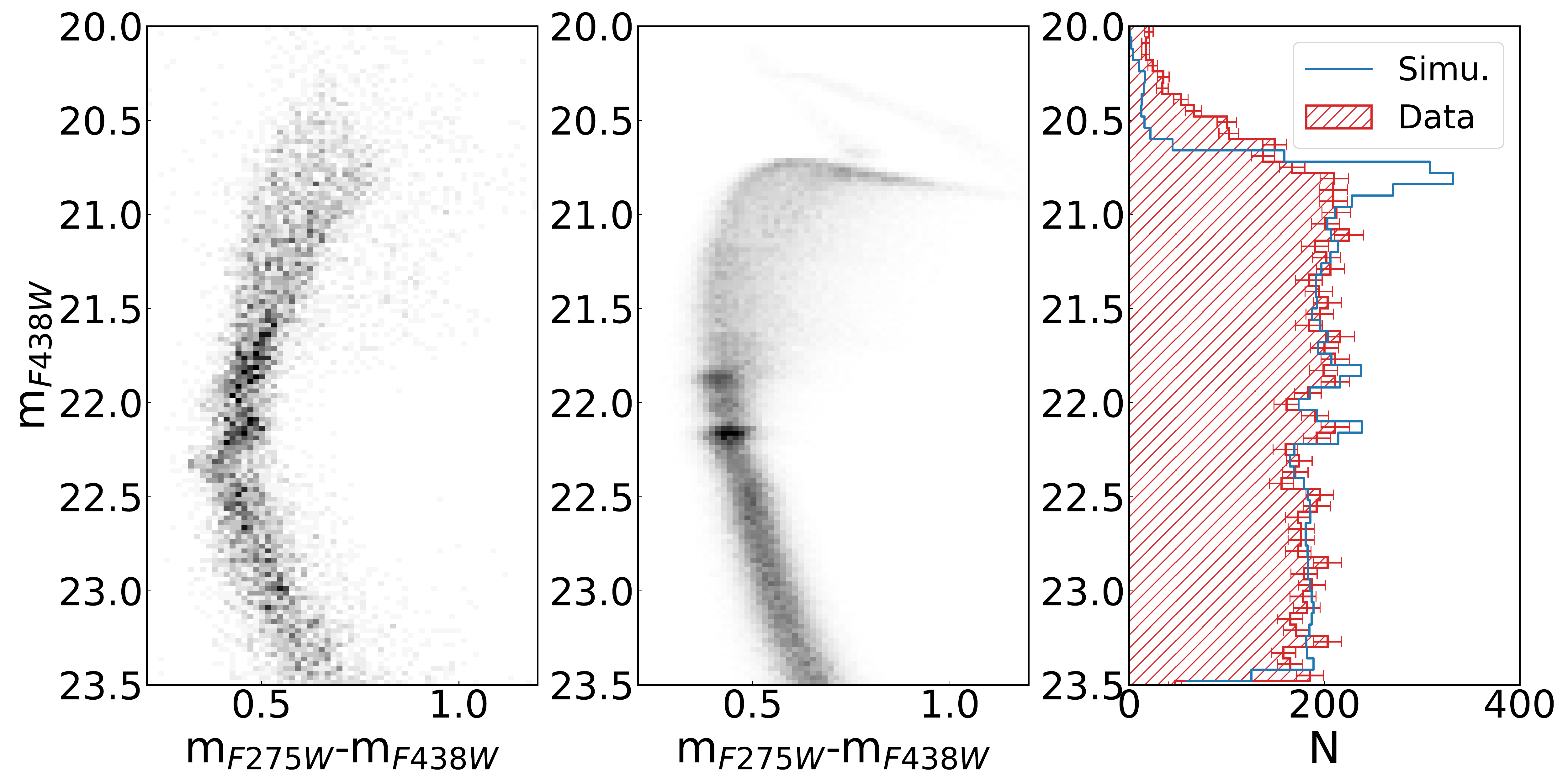}
\caption{The Hess diagram of the photometric data (left panel) and of the simulation (central panel) in the m$_{\rm F438W}$ versus m$_{\rm F275W}$--m$_{\rm F438W}$ plane. In the right panel we report the comparison of the histograms of stellar counts as a function of
m$_{\rm F438W}$ between the data (red) and the simulation (blue).}
\label{fig12simuv}
\end{figure}

\begin{figure}
\centering
\includegraphics[width=8.5cm]{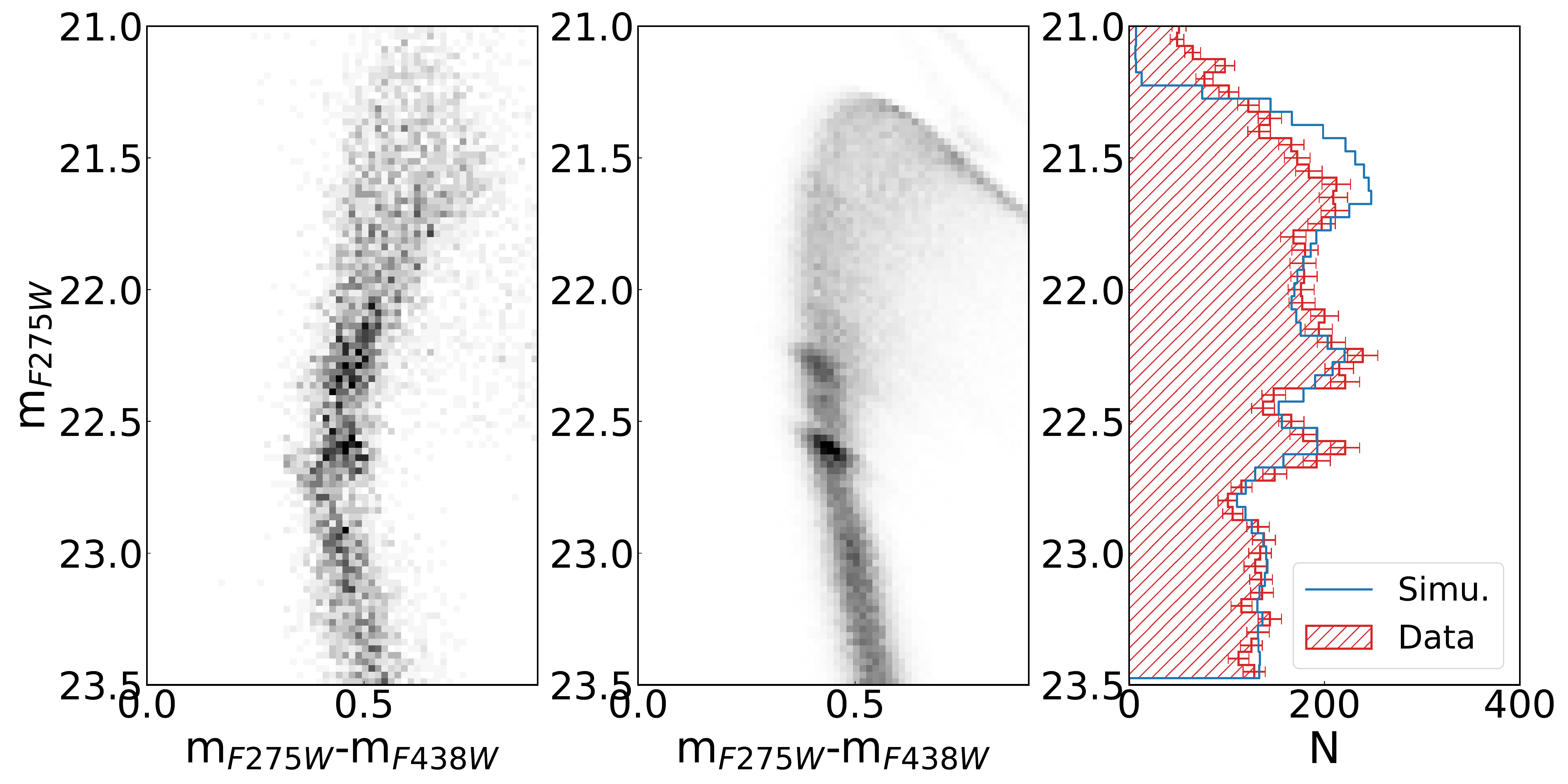}
\caption{The Hess diagram of the photometric data (left panel) and of the simulation (central panel) in the m$_{\rm F275W}$ versus m$_{\rm F275W}$--m$_{\rm F438W}$ plane.In the right panel we report the comparison of the histograms of stellar counts as a function of
m$_{\rm F275W}$ between the data (red) and the simulation (blue). 
}
\label{fig13simuv}
\end{figure}

\begin{figure}
\centering
\includegraphics[width=8.5cm]{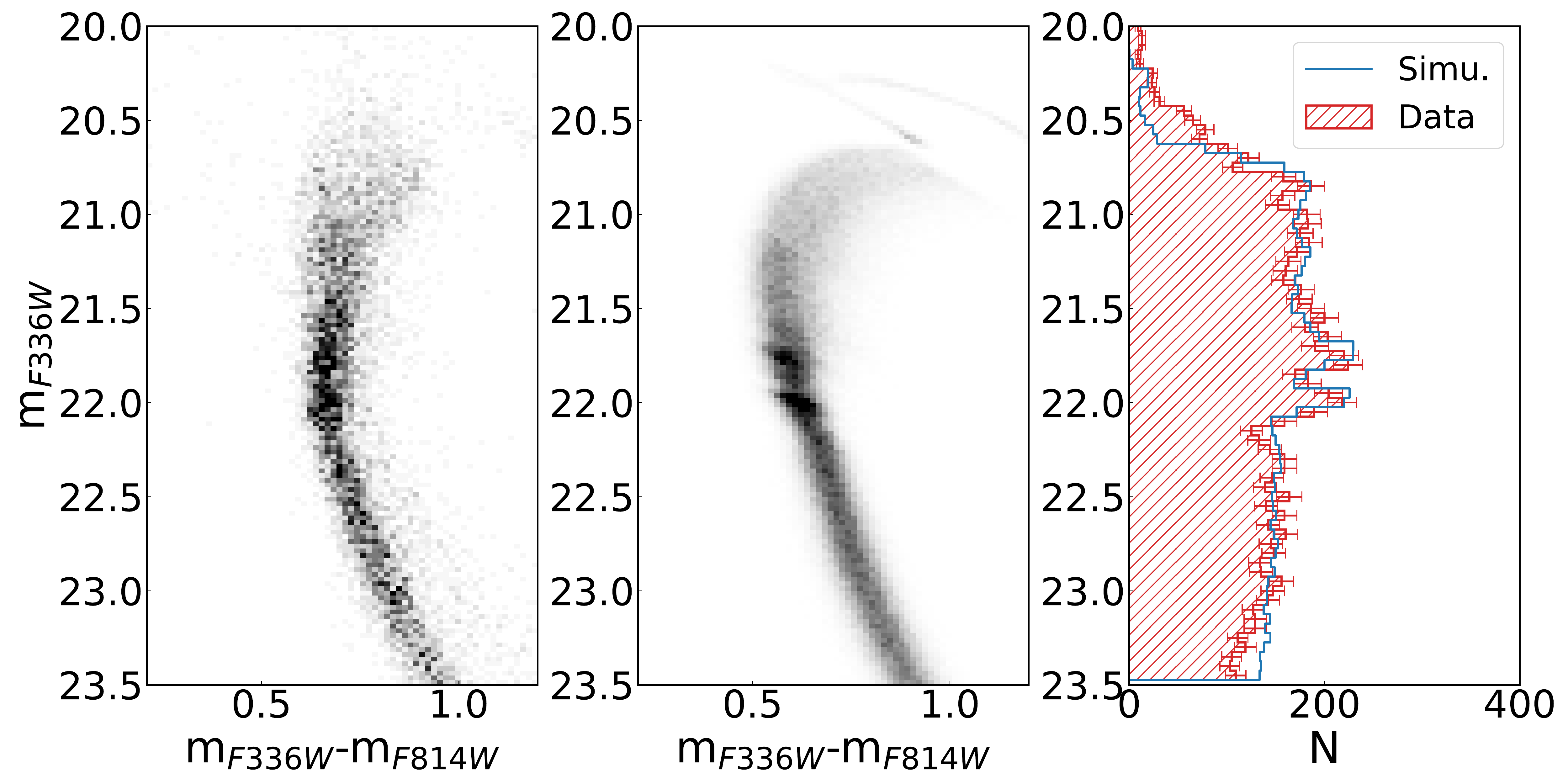}
\caption{The Hess diagram of the photometric data (left panel) and of the simulation (central panel) in the m$_{\rm F336W}$ versus m$_{\rm F336W}$--m$_{\rm F814W}$ plane. In the right panel we report the comparison of the histograms of stellar counts as a function of
m$_{\rm F336W}$ between the data (red) and the simulation (blue).}
\label{fig14simopt}
\end{figure}

\begin{figure}
\centering
\includegraphics[width=8.5cm]{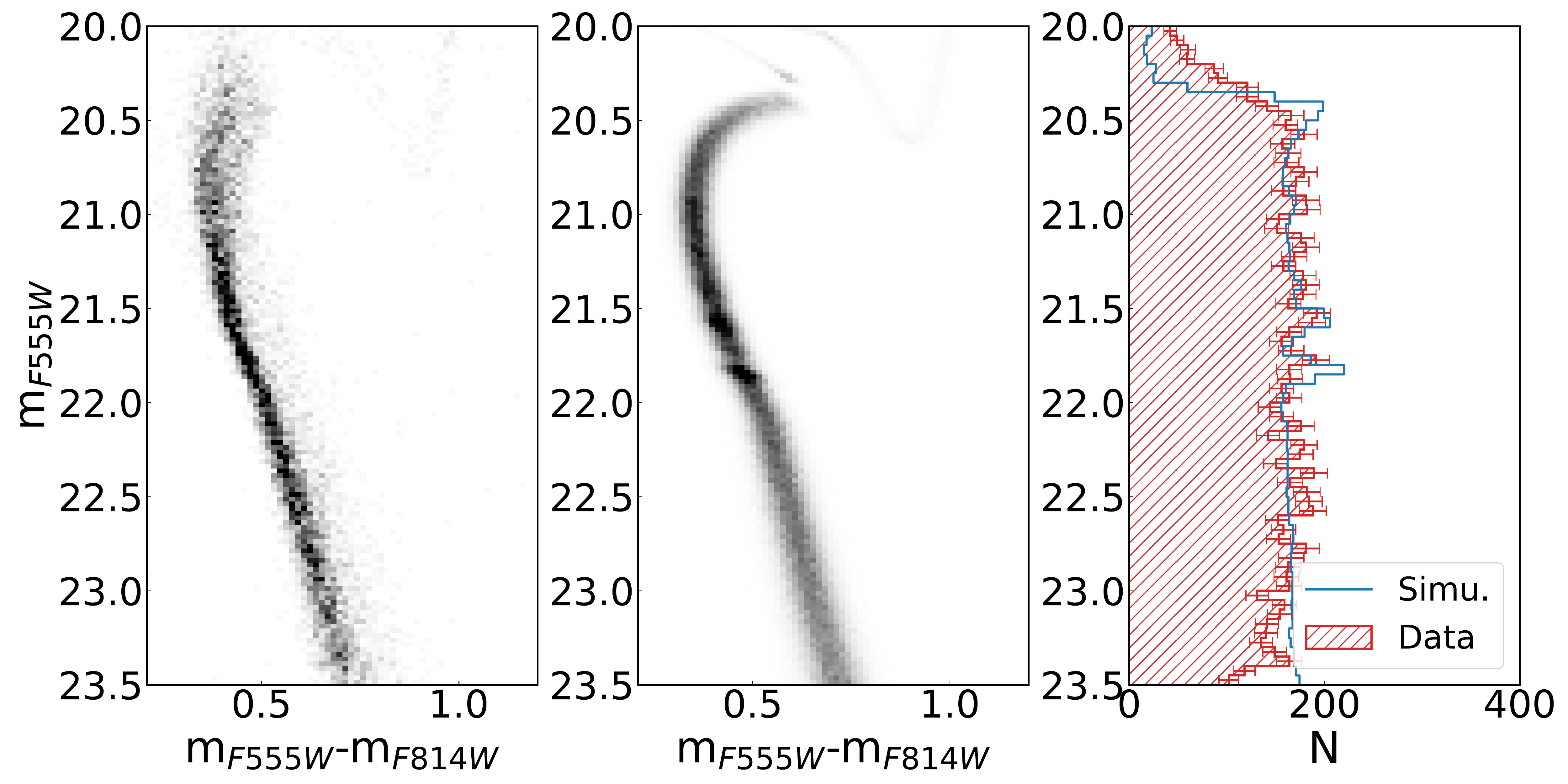}
\caption{The Hess diagram of the photometric data (left panel) and of the simulation (central panel) in the m$_{\rm F555W}$ versus m$_{\rm F555W}$--m$_{\rm F814W}$ plane. In the right panel we report the comparison of the histograms of stellar counts as a function of
m$_{\rm F555W}$ between the data (red) and the simulation (blue).}
\label{fig15simopt}
\end{figure}

In Fig. \ref{fig12simuv}-\ref{fig13simuv}-\ref{fig14simopt}-\ref{fig15simopt}, we report the comparison between NGC\,1783 data (left panel) and the results from the simulation which best reproduce the distribution of the stars observed (central panel). The m$_{\rm F438W}$ versus m$_{\rm F275W}$--m$_{\rm F438W}$ CMD reported in Fig. \ref{fig12simuv}
is the best plane where we can calibrate the distribution of the optical depth necessary to reproduce the UV-dim stars. We select $\sigma=0.045$ for the gaussian distribution of $\tau_{10}$. The mean value $\tau_{av}$ for stars in the brightest part of the eMSTO is $\tau_{av}$=0.08, for stars with $1.4<\rm M_{\rm in}/M_{\odot}<1.6$ evolving in the range of magnitudes $20.6<\rm m_{\rm F438W}<21.4$. The mean $\tau_{av}$\ and $\sigma$ decreases for stars evolving at fainter magnitudes until $\tau_{av}$=0.015 and $\sigma=0.25$ in correspondence to the peaks/gaps. 

In Fig. \ref{fig14simopt} and \ref{fig15simopt} we notice how the absorption of dust gradually decreases when considering longer wavelenghts magnitudes; indeed UV-dim stars are mixed to the rest of the MS stars in the m$_{\rm F336W}$ versus m$_{\rm F336W}$--m$_{\rm F814W}$ and m$_{\rm F555W}$ versus m$_{\rm F555W}$--m$_{\rm F814W}$ planes, both in the data and in the simulation.
In the right panel of Fig. \ref{fig12simuv}-\ref{fig13simuv}-\ref{fig14simopt}-\ref{fig15simopt}, we show the comparison of the histograms of stellar counts as a function of the CMD corresponding magnitude between the data (red) and the simulation (blue). 

\section{Discussion and Conclusions}
\label{discussion}
The presence of UV-dim stars at the level of the eMSTO in NGC\,1783 has been modelled by assuming that these stars are affected by circumstellar dust, whose absorption is strongly wavelength dependent. As the dust is probably distributed in a disk or ring around the stars, a  necessary consequence of this model is that stars affected by dust but seen at low inclination angles will also populate the eMSTO region. 

Although improvement in the comparison between simulations and data requires isochrones matching the eMSTO region better than those available today, using of a selection of the stars located at the bluer edge of the eMSTO and reasonable approximations on the dust effect are sufficient to gain considerable insight into the consequences of this approach.
{\it Our analysis shows that the entire eMSTO can be explained by dust absorption.} The proposed model represents a major shift in the interpretation of the eMSTO: the location of stars within the turnoff fan depends on the effect of circumstellar dust on the intrinsic colors, modulated by the inclination angle of view of the rotation axis, and the stars mostly affected by dust will be redder.
The same high inclination angle will cause the projected rotation velocity $v \sin$\, i to be maximum in the reddest stars. As the more rapidly rotating stars will be more subject to the mass loss events causing the presence of circumstellar dust, the direct correlation often found between the projected rotation rate and the redder color finds a direct simple explanation.

This model thus naturally implies the formation of an extended turnoff, independently from the inclusion of other factors (such as a difference in age, or different rotational evolutionary paths). 
 The role of dust bypasses the necessity to understand whether only the very rapidly rotating stars, close to break-up rotation ($\omega/\omega_{\rm crit} \sim 0.9$) are flattened enough that limb and gravity darkening can result in appreciably different stellar \teff\ and emission according to the viewing angle, as in Geneva models, or the effect is relevant also at lower $\omega/\omega_{\rm crit}$ (e.g. $ \sim 0.6$, as in MIST models). In any case, the faster spinning stars are expected to be more subject to mass loss and dust formation, and thus to be redder. It will be very important to check the role of this effect in other clusters and understand whether it is the main reason for the presence of the eMSTO also in younger clusters having a split main sequence \citep[e.g.][]{milone2016,milone2018,kamann2023}, where the rotational interpretation was well accepted. 

A further argument in favour of the relation between the dust and the fan shape is  that these clusters mostly show a subgiant branch thickness much tighter than expected if the isochrone location is dependent on the rotation rate, as described by the models including either rotation, as e.g. the MIST models ---see  Fig.\,2 in \citet{kamann2020}), or different extension of the convective cores, as in the cloud isochrones by \cite{johnston2019}. 

The two peaks/gaps in the density distribution of the upper MS in NGC\,1783 are probaby linked to the transition between stars which lose mass to an excretion disk and stars having convective envelopes and therefore braked rotation and no mass loss. A simple treatment of the mass loss problem based naively on a single  isochrone shows that a mass loss of a few $\sim 10^{-3}$\Msun, cumulated in episodes occurring along the MS lifetime may reproduce the peaks, although also the onset of dust absorption plays some role in defining the upper peak. 

Although this is a first simple attempt to simulate a complex situation, we suggest that both the zig-zag MS, the shape of the eMSTO and the UV--dim stars are a consequence of the mass loss from rapidly rotating modes.

The phenomenon of extended MS turnoffs has been easily discovered in the rich MC clusters, but it has been shown to occur also in the much less populated Galactic open clusters \citep[e.g.][]{marino2018b}. Also the original B\"ohm-Vitense gap at $\sim$7200\,K and the dimmer Hyades gap at 6800\,K have been discovered thanks to Galactic stellar observations. We dare to suggest that both these gaps are not related only to the original explanations, but are also due to the transition from rapidly rotating stars suffering mass loss during their MS lifetime to non rotating stars.

In conclusion, while the study of the eMSTO is still to be fully exploited,  the additional role played by dust is a critical ingredient to be taken into account, and is opening a new window in our understanding of stellar evolution.

\section{Data availability}
The data underlying this article will be shared on reasonable request to the corresponding author.

\section*{Acknowledgements}
We thank Antonella Natta and Andrea Dupree for useful discussions on the role of circumstellar dust. MT acknowledges support from the ERC Consolidator Grant funding scheme (project ASTEROCHRONOMETRY, https://www.asterochronometry.eu, G.A. n. 772293). FD and PV ackwnoledge support from the INAF-GTO-Grant 2022 entitled "Understanding the formation of globular clusters with their multiple stellar generations" (P.I. A. Marino). 



\bibliographystyle{mnras}
\bibliography{lmc}




\bsp    
\label{lastpage}
\end{document}